\documentclass{ar-1col}
\usepackage[numbers]{natbib}
\usepackage{subfigure}
\jname{Annu. Rev. Cond. Matter Phys.}\jvol{8}\jyear{2017}
\doi{10.1146/(please add article doi)}

\begin{document}

\markboth{Uwe C. T\"auber}
	{Phase Transitions and Scaling in Systems Far From Equilibrium}
\title{Phase Transitions and Scaling in Systems Far From Equilibrium}

\author{Uwe C. T\"auber
\affil{Department of Physics and Center for Soft Matter and Biological Physics \\
      (MC 0435), Virginia Tech, Blacksburg, VA 24061, USA; email: tauber@vt.edu}}

\begin{abstract} 
Scaling ideas and renormalization group approaches proved crucial for a deep
understanding and classification of critical phenomena in thermal equilibrium. 
Over the past decades, these powerful conceptual and mathematical tools were 
extended to continuous phase transitions separating distinct non-equilibrium 
stationary states in driven classical and quantum systems. In concordance with 
detailed numerical simulations and laboratory experiments, several prominent 
dynamical universality classes have emerged that govern large-scale, long-time 
scaling properties both near and far from thermal equilibrium. These pertain to 
genuine specific critical points as well as entire parameter space regions for 
steady states that display generic scale invariance. The exploration of 
non-stationary relaxation properties and associated physical aging scaling 
constitutes a complementary potent means to characterize cooperative dynamics in 
complex out-of-equilibrium systems. This article describes dynamic scaling
features through paradigmatic examples that include near-equilibrium critical 
dynamics, driven lattice gases and growing interfaces, correlation-dominated
reaction-diffusion systems, and basic epidemic models.
\end{abstract}



\maketitle



\section{INTRODUCTION}

The fiercely idealized and simplifying notion of thermal equilibrium that treats
systems devoid of contact with their environment in a fully relaxed state long 
after their preparation has nevertheless had profound impact in setting a 
framework for a microscopic basis of phenomenological thermodynamics and our 
understanding of macroscopic condensed matter systems built from many interacting 
constituents.
Most dynamic processes in nature however occur in out-of-equilibrium settings, 
where the system under consideration is either subject to strong time-dependent
external perturbations (beyond the linear response regime), or where, e.g., a 
non-vanishing energy, mass, or electric current flows through it.
Despite considerable effort, a fundamental conceptual framework of non-equilibrium
systems akin to equilibrium statistical mechanics is still lacking.
This is even true for non-equilibrium steady states, whose macroscopic observables 
are time-independent:
Neither do we have a general recipe to construct the associated probability
distributions, nor to generically characterize the stationary probability currents 
that are indispensable for their complete classification.

A promising avenue to achieve partial progress in this important and difficult 
area is naturally provided by the study of continuous phase transitions and the
accompanying critical phenomena, since these should be governed by universal
features that could hopefully be amenable to systematic characterization through
dynamical universality classes within a renormalization group methodology.
Indeed, numerical and theoretical investigations of non-equilibrium phase 
transitions in classical stochastic dynamical systems have over the past four 
decades led to notable advances; and this brief review attempts to summarize 
some of the key results.
In addition, it has transpired that as compared with thermal equilibrium, generic 
scale invariance represents a far more ubiquitous feature in driven systems.
Over only the past few years, detailed experiments have unambiguously confirmed 
the relevance of the most prominent non-equilibrium universality classes beyond
the realms of mathematics and computer simulation, and quantitatively checked the
predicted power laws.
In the quantum world too, both theoretical and experimental studies of externally
driven systems as well as of the non-equilibrium relaxation kinetics following 
sudden parameter quenches have recently become a highly active research field. 
One would therefore anticipate that the analysis of strong spatio-temporal
fluctuations and long-range correlations with associated scaling phenomena is
likely to remain important in condensed matter physics and materials science, but 
will gain prominence in (bio-)chemistry, systems biology, ecology, and finance.

\section{CRITICAL SCALING IN THERMAL EQUILIBRIUM}

To set the stage, we begin with a brief review of scaling theory as devised for 
continuous phase transitions and critical points in thermal equilibrium.

\subsection{Thermodynamic Singularities at Critical Points}

Continuous or second-order phase transitions in thermal equilibrium are 
characterized by the emergence of characteristic thermodynamic singularities in
the vicinity of the critical point \cite{Fisher67, Stanley71, Ma76, Yeomans91, 
Goldenfeld92, Binney93, Chaikin95, Cardy96}.
These appear specifically in the thermodynamic limit, where the number of
constituents $N$ along with other extensive quantities (such as volume $V$, 
free energy $F$, entropy $S$, etc.) are taken to infinity, with the respective
densities $v = V / N$, $f = F / N$, $s = S / N \ldots$ held fixed.
Typically, there are two relevant control parameters that govern the behavior of
the singular contributions of the free energy $f_s$ per particle in the vicinity 
of the critical point: the relative distance $\tau = (T - T_c) / T_c$ from the 
critical temperature $T_c$, and the magnitude of a (symmetry-breaking) external 
field $h$, thermodynamically conjugate to the order parameter $\phi$ that 
characterizes the phase transition. 
The critical point is then located at $\tau = 0$, $h = 0$, and the scaling
hypothesis for $f_s$ asserts that it assumes the form of a generalized homogeneous
function as $\tau \to 0$, $h \to 0$:
\begin{equation}
  f_s(\tau,h) = |\tau|^{2 - \alpha} \, {\hat f}_\pm\bigl( h / |\tau|^\Delta \bigr)
  \ .
\label{fschyp}
\end{equation}
The singular contribution to the free energy density hence satisfies a remarkable
two-parameter scaling law, with distinct analytic scaling functions 
${\hat f}_\pm(y)$ for $T > T_c$ and $T < T_c$, respectively, which only depend on
the ratio $y = h / |\tau|^\Delta$, and satisfy ${\hat f}_\pm(0) =$ const., and 
with merely two independent thermodynamic critical exponents $\Delta > 0$ and 
$\alpha$.

These are related to the singular behavior of certain thermodynamic quantities
near the critical point:
The specific heat at $h = 0$ scales according to 
$C_{h = 0} \sim - \left( \partial^2 f_s / \partial \tau^2 \right)_{h = 0} 
 = C_\pm |\tau|^{- \alpha}$, indicating a divergence at $T_c$ if $\alpha > 0$, and 
a cusp singularity for $\alpha < 0$.
The order parameter equation of state is obtained via
$\phi(\tau,h) = - \left( \partial f_s / \partial h \right)_\tau = 
 - |\tau|^{2 - \alpha - \Delta} \, {\hat f}_\pm'\bigl( h / |\tau|^\Delta \bigr)$,
which yields the coexistence curve in the low-temperature ordered phase 
($\tau < 0$) at $h = 0$: $\phi(\tau,0) \sim |\tau|^\beta$, where 
$\beta = 2 - \alpha - \Delta$.
Next, the critical isotherm at $\tau = 0$ follows from the requirement that the 
$\tau$ dependence in ${\hat f}_\pm'$ must cancel the singular prefactor:  
${\hat f}_\pm'(y \to \infty) \sim y^{(2 - \alpha - \Delta)/ \Delta}$,
and consequently $\phi(0,h) \sim h^{1 / \delta}$ with $\delta = \Delta / \beta$.
Finally, the isothermal order parameter susceptibility becomes $\chi_\tau \sim 
 \left( \partial \phi / \partial h \right)_{\tau, \, h = 0} = \chi_\pm 
 |\tau|^{- \gamma}$, where $\gamma = \alpha + 2 \left( \Delta - 1 \right)$ (on 
both sides of the phase transition).
Eliminating $\Delta = \beta \delta$ gives the following set of scaling relations
that link the various thermodynamic critical exponents:
\begin{equation}
  \alpha + \beta \left( 1 + \delta \right) = \alpha + 2 \beta + \gamma = 2 \ , 
  \quad \gamma = \beta \left( \delta - 1 \right) \, . 
\label{screls}  
\end{equation}

Landau's general mean-field description of phase transitions relies on an 
expansion of the free energy density $f_s$ in terms of the order parameter $\phi$,
subject to the fundamental symmetries of the physical system under consideration.
For example, for a scalar order parameter $\phi$ with discrete inversion ($Z_2$) 
symmetry $\phi \leftrightarrow - \phi$ one would expand $f_s$ as follows:
\begin{equation}
  f_s(\phi) = \frac{r}{2}\, \phi^2 + \frac{u}{4!} \, \phi^4 + \ldots - h \, \phi 
  \ .
\label{landex}
\end{equation}
For $u > 0$, a continuous phase transition ensues at $r = 0$, where the 
spontaneous order parameter changes from $\phi_0 = 0$ at $r > 0$ to either of the
two degenerate values $\phi_\pm = \pm \sqrt{6 |r| / u}$ for $r < 0$, whence 
$r = a \left( T - T_c^0 \right)$ with $a > 0$, and where $T_c^0$ denotes the 
mean-field critical temperature.
The corresponding mean-field critical exponents in Landau theory are readily found
to be $\alpha = 0$, $\beta = 1/2$, $\gamma = 1$, $\delta = 3$, and $\Delta = 3/2$.

\subsection{The Role of Spatial Fluctuations near Continuous Phase Transitions}

The divergence of the order parameter susceptibility, which according to the
equilibrium fluctuation-response theorem is intimately related to the mean-square
order parameter fluctuations $\chi_T \sim \left( \langle \phi^2 \rangle - 
 \langle \phi \rangle^2 \right) / k_{\rm B} T$, indicates that the latter become
very prominent in the vicinity of the critical point.
Hence spatial fluctuations need to be properly included in the theoretical
description of continuous phase transitions \cite{Fisher67, Stanley71, Ma76, 
Amit84, Itzykson89, Yeomans91, Goldenfeld92, Binney93, Zinn93, Chaikin95, 
Cardy96, Vasiliev04, Tauber14}.
To this end, one may generalize the Landau expansion \ref{landex}. to 
inhomogeneous order parameter configurations $S({\vec x})$ through a 
coarse-grained effective Landau--Ginzburg--Wilson Hamiltonian (in $d$ space
dimensions)
\begin{equation}
  \frac{{\cal H}[S]}{k_{\rm B} T} = \int \! d^dx \left[ \, \frac{r}{2} \, 
  S({\vec x})^2 + \frac12 \left[ {\vec \nabla} S({\vec x}) \right]^2 
  + \frac{u}{4 !} \, S({\vec x})^4 - h({\vec x}) \, S({\vec x}) \, \right] \, ,
\label{lgwham}
\end{equation}
where $h({\vec x})$ represents a local external field.
Here we assume spatial inhomogeneities to be energetically unfavorable, and have
absorbed the positive coefficient for the gradient term 
$\propto \left[ {\vec \nabla} S({\vec x}) \right]^2$  into the scalar order 
parameter field $S({\vec x})$.
Within the canonical framework of statistical mechanics, the probability density 
for any specific configuration $S({\vec x})$ is given by the Boltzmann factor
${\cal P}_s[S] = \exp \left( - {\cal H}[S] / k_{\rm B} T \right) / {\cal Z}[h]$. 
The partition function 
${\cal Z}[h] = \int \! {\cal D}[S] \ e^{- {\cal H}[S] / k_{\rm B} T}$ and any 
expectation values of physical observables $A[S]$ such as 
$\phi = \langle S({\vec x}) \rangle$ are represented through functional integrals:
$\langle A[S] \rangle = \int \! {\cal D}[S] \, A[S({\vec x})] \, {\cal P}_s[S]$.

Near the critical point, order parameter fluctuations become strong and 
long-range, which is encoded in the asymptotic divergence of the characteristic
correlation length $\xi = \xi_\pm \, |\tau|^{- \nu}$ as $|\tau| \to 0$ with 
$\nu > 0$.
The scaling hypothesis for the two-point order parameter correlation function
$C({\vec x}) = \langle S({\vec x}) \, S(0) \rangle - \phi^2$ then asserts that
\begin{equation}
  C(\tau,{\vec x}) = |{\vec x}|^{- (d - 2 + \eta)} \, 
  {\widetilde C}_\pm({\vec x} / \xi) \ ,
\label{spcorr}
\end{equation}
which defines Fisher's anomalous correlation exponent $\eta$.
Away from the critical regime, typically ${\widetilde C}_\pm(y) \sim e^{-y}$
decays exponentially, while at criticality ${\widetilde C}_\pm(0) =$ const., i.e.,
$C({\vec x})$ falls off only algebraically at large distances 
$|{\vec x}| \to \infty$. 
In this limit, one expects 
$\langle S({\vec x}) \, S(0) \rangle \to \phi^2 \sim (- \tau)^{2 \beta}$ for
$\tau < 0$, whence we identify $2 \beta = \nu \left( d - 2 + \eta \right)$.
For the spatially Fourier--transformed correlation function \ref{spcorr}. implies
\begin{equation} 
  C(\tau,{\vec q}) = \int \frac{d^dq}{(2 \pi)^d} \ C(\tau,{\vec x}) \,
  e^{-i {\vec q} \cdot {\vec x}} 
  = |{\vec q}\,|^{- 2 + \eta} \, {\hat C}_\pm({\vec q} \, \xi) \, ,
\label{corsch}
\end{equation}
with ${\hat C}(y \to \infty) \to$ const., and consequently the thermodynamic 
susceptibility follows as
\begin{equation}
  \chi(\tau,{\vec q} = 0) \sim C(\tau,{\vec q} = 0) \sim \xi^{2 - \eta} 
  \sim |\tau|^{- \gamma} \ , \quad \gamma = \nu \left( 2 - \eta \right) \, ,
\label{screld}
\end{equation}
providing us with a second another scaling relation that connects the 
thermodynamic critical exponents with $\eta$ and $\nu$.
Crucially, the above scaling analysis explains the thermodynamic critical point 
singularities to be induced by the diverging spatial correlations near a 
continuous phase transition, and combining \ref{screls}. and \ref{screld}. yields
the hyperscaling relations
\begin{equation}
  \alpha = 2 - d \, \nu \ , \quad 
  \beta = \frac{\nu}{2} \left( d - 2 + \eta \right) \, , \quad 
  \delta = \frac{d + 2 - \eta}{d - 2 + \eta}
\label{hypscr}
\end{equation}
that contain the spatial dimension $d$.

Eqs.~\ref{hypscr}. hold below the upper critical dimension, which for generic 
continuous equilibrium phase transitions governed by effective free energy 
functionals of the form \ref{lgwham}. turns out to be $d_c = 4$.
This can be readily inferred from the following direct dimensional analysis, but
put on firm grounds through a systematic renormalization group treatment
\cite{Ma76, Amit84, Itzykson89, Goldenfeld92, Binney93, Zinn93, Cardy96, 
Vasiliev04, Tauber14}:
Since ${\cal H}[S] / k_{\rm B} T$ is dimensionless, we infer that in terms of the
wave vector or inverse length scale $[{\vec q}] = [{\vec x}]^{-1} = \mu$, the 
fluctuating order parameter field scales as $[S({\vec x})] = \mu^{(d - 2) / 2}$.
Thus we find $[r] = \mu^2$ and $[h({\vec x})] = \mu^{(d + 2) / 2}$ with positive
scaling dimensions, which indicates that these two basic external control 
parameters constitute relevant scaling fields in the renormalization group sense.
For the non-linear coupling one obtains the scaling dimension $[u] = \mu^{4 - d}$;
this combines with the temperature variable $r$ to an effective dimensionless
coupling $u \, |r|^{(d - 4) / 2}$, which scales toward zero near the phase 
transition at $r = 0$ in dimensions $d > 4$, and hence is irrelevant.
The hyperscaling relations \ref{hypscr}. hold if $d$ is replaced with $d_c = 4$.
In high dimensions therefore the mean-field or Gaussian critical exponents 
$\eta = 0$ and $\nu = 1/2$, essentially obtained by setting $u = 0$ in the 
Hamiltonian \ref{lgwham}., correctly describe the universal critical scaling 
behavior.
In contrast, the effective non-linear coupling diverges as $|r| \to 0$ for 
$d < d_c = 4$, signifying the crucial importance of critical fluctuations on
the thermodynamics of continuous phase transitions. 
Here, the scaling exponents become modified; e.g., the correlation exponents 
$\eta$, $\nu$, and hence the values of $\beta$ and $\gamma$ are enhanced, whereas 
$\delta$ is reduced relative to the mean-field predictions.

Yet these critical exponents, along with certain amplitude ratios such as 
$C_+ / C_-$, $\chi_+ / \chi_-$, and $\xi_+ / \xi_-$, and even the full scaling 
functions ${\hat f}_\pm$ and ${\hat C}_\pm$ remain universal features that
characterize continuous phase transitions according to the system's fundamental 
symmetries, and thus depend only on the spatial dimension $d$, number of order
parameter components $n$, and if applicable the power law of involved long-range
interactions.
In the renormalization group approach, these broad equivalence or universality 
classes are defined through infrared-stable fixed points for the associated
renormalization group flows that describe the scale dependence of running 
couplings.
Thus emerge identical critical scaling properties for diverse systems with 
distinct microscopic interactions, defined on different lattices or a continuum; 
for example, the critical points in Ising magnets and for liquid-gas transitions 
are described by the free energy functional \ref{lgwham}. for a scalar order 
parameter.

\section{Near-Equilibrium Critical Dynamics}

Next we explore the extension of scaling theory for equilibrium thermodynamics and
correlation functions to dynamic phenomena near critical points \cite{Ferrell67, 
Ferrell68, Halperin69, Hohenberg77, Chaikin95, Vasiliev04, Folk06, Tauber14}.

\subsection{Dynamic Scaling Hypothesis and Relaxational Critical Dynamics}

As spatially correlated regions grow tremendously upon approaching a continuous 
phase transition, the characteristic relaxation time $t_c$ associated with the 
order parameter kinetics should increase dramatically as well, 
$t_c(\tau) \sim \xi(\tau)^z \sim |\tau|^{- z \nu}$.
This critical slowing-down is described by a dynamic critical exponent 
$z = \nu_t / \nu$, which may be interpreted as the ratio of correlation exponents
along the temporal and spatial directions. 
One may thus formulate a dynamic scaling ansatz for the corresponding 
wavevector-dependent typical frequency scale,
\begin{equation}
  \omega_c(\tau,{\vec q}) = |{\vec q}\,|^z \, {\hat \omega}_\pm({\vec q} \, \xi) 
  \ ,
\label{dynscf}
\end{equation}
with ${\hat \omega}_\pm(y \to \infty) \to$ const. 
This implies a critical dispersion relation 
$\omega_c(0,{\vec q}) \sim |{\vec q}\,|^z$.

In thermal equilibrium, the dynamical response and correlation functions are
intimately connected through the fluctuation-dissipation theorem \cite{Forster83, 
Lovesey86, Kubo91, Chaikin95, VanVliet10, Tauber14}
\begin{equation}
  C(\tau,{\vec q},\omega) 
  = \int \frac{d\omega}{2 \pi} \ C(\tau,{\vec q},t) \, e^{i \omega t} 
  = \frac{2 k_{\rm B} T}{\omega} \ {\rm Im} \, \chi(\tau,{\vec q} , \omega) \ .
\label{fldist}
\end{equation}
The dynamic scaling hypotheses for the asymptotic critical properties of the 
time-dependent correlation function and dynamical susceptibility then become
\begin{equation}
  C(\tau,{\vec x},t) = |{\vec x}|^{-(d - 2 + \eta)} \, 
  {\widetilde C}_\pm({\vec x} / \xi , t / t_c) \ , \quad
  \chi(\tau,{\vec q},\omega) = |{\vec q}\,|^{- 2 + \eta} \, 
  {\hat \chi}_\pm({\vec q} \, \xi , \omega \, t_c) \ , 
\label{dynscr}  
\end{equation}
which generalize the static scaling laws \ref{spcorr}., \ref{corsch}.
As a consequence of the stringent constraints imposed by the 
fluctuation-dissipation relation \ref{fldist}., the very same three independent 
critical exponents $\nu$, $\eta$, and $z$ fully characterize the universal scaling
regimes in Eqs.~\ref{dynscr}.
Appropriate dynamical scaling variants can also be postulated for transport 
coefficients.

The thermodynamics of genuine quantum phase transitions located at zero 
temperature follows essentially the same scaling phenomenology 
\cite{Continentino94, Vojta03, Sachdev11, Tauber14}.
By means of the coherent-state path integral formalism \cite{Negele88, Tauber14}, 
quantum fluctuations are encapsulated through an imaginary-time integration in 
addition to the $d$-dimensional spatial integral in the corresponding effective 
action, or over Matsubara frequencies $\omega$ in its Fourier representation.
The quantum counterpart to a classical coarse-grained critical Hamiltonian such
as \ref{lgwham}. is thus inherently dynamic in nature, or equivalently constitutes
a $(d + 1)$-dimensional anisotropic field theory with anisotropy exponent $z$.
Correspondingly, one needs to replace the space dimension $d$ with $d + z$ in the 
associated hyperscaling relations \ref{hypscr}.

The mathematical description of near-equilibrium critical dynamics (at $T > 0$) 
explicitly exploits the critical slowing-down of the order parameter fluctuations.
The ensuing time scale separation with respect to most other dynamical degrees of
freedom affords an effective representation through stochastic Langevin equations,
wherein all fast modes are projected onto fast thermal noise.
In the simplest scenario, one imposes purely relaxational kinetics of the order
parameter field $S({\vec x},t)$ towards a minimum of the free energy functional
\ref{lgwham}., 
\begin{equation} 
  \frac{\partial S({\vec x},t)}{\partial t} = - D \, 
  \frac{\delta {\cal H}[S]}{\delta S({\vec x},t)} + \zeta({\vec x},t) \ ,
\label{relmda}
\end{equation}
with relaxation rate $D$ and uncorrelated (white) Gaussian random noise $\zeta$ 
that is completely characterized by its vanishing mean and variance:
\begin{equation}
  \big\langle \zeta({\vec x},t) \big\rangle = 0 \ , \quad 
  \left\langle \zeta({\vec x},t) \, \zeta({\vec x\,}',t') \right\rangle 
  = 2 \Gamma \, \delta({\vec x} - {\vec x\,}') \, \delta(t - t') \ . 
\label{stnmda}
\end{equation} 
As can be inferred from the associated Fokker--Planck equation, the time-dependent
configurational probability distribution ${\cal P}[S,t]$ asymptotically approaches
the stationary canonical distribution ${\cal P}[S,t \to \infty] \to {\cal P}_s[S]$
provided the Einstein relation $\Gamma = D \, k_{\rm B} T$ is imposed that links
the noise strength $\Gamma$ with the relaxation rate $D$ and temperature $T$
\cite{Risken84, Chaikin95, Schwabl06, VanVliet10, Tauber14}.  

In Hohenberg and Halperin's alphabetical classification, the critical relaxational
dynamics of a non-conserved order parameter \ref{relmda}. with noise \ref{stnmda}.
is labeled model A \cite{Hohenberg77}.
It is straightforward to infer the dynamic critical exponent $z = 2$ for model A 
in the Gaussian approximation ($u = 0$).
Both Monte Carlo simulations and renormalization group analysis yield that for
$d < d_c = 4$ fluctuations slightly enhance this value.
The result is usually parametrized as $z = 2 + c \, \eta$, where, e.g., 
$c = 6 \ln (4/3) - 1 + O(\epsilon) > 0$ to second order in the dimensional 
expansion with respect to $\epsilon = 4 - d$ \cite{Hohenberg77, Folk06, Tauber14}.
If in contrast the order parameter is conserved, and its local density hence 
satisfies a continuity equation, its spatial fluctuations must relax diffusively.
Consequently, the constant relaxation rate $D$ is to be replaced with the 
diffusion operator $- D \, {\vec \nabla}^2$ in the Langevin equation \ref{relmda}.
as well as in the noise correlator \ref{stnmda}. for this so-called model B.
The order parameter conservation law moreover enforces the exact exponent scaling 
relation $z = 4 - \eta$.

\subsection{Critical Dynamics with Reversible Couplings to Other Conserved Modes}

Aside from the order parameter constituting a conserved or non-conserved quantity,
a further splitting of the static critical behavior into distinct dynamical 
universality classes in thermal equilibrium is caused by the presence of other
conserved densities and the possible coupling of these additional slow modes to 
the order parameter fluctuations \cite{Ma76, Hohenberg77, Goldenfeld92, Chaikin95,
Vasiliev04, Folk06, Tauber14}.
Such situations can be described by a set of coupled Langevin equations for 
coarse-grained mesoscopic stochastic variables $S^\alpha$ of the general form  
\begin{eqnarray} 
  &&\frac{\partial S^\alpha({\vec x},t)}{\partial t} = F^\alpha[S]({\vec x},t) 
  + \zeta^\alpha({\vec x},t) \ ,
\label{langen} \\ 
  &&\big\langle \zeta^\alpha({\vec x},t) \, \zeta^\beta({\vec x\,}',t') 
  \big\rangle = 2 L^\alpha[S] \, \delta({\vec x} - {\vec x\,}') \, \delta(t - t')
  \, \delta^{\alpha \beta} \ .
\label{noigen}
\end{eqnarray}
Naturally $\langle \zeta^\alpha({\vec x},t) \rangle = 0$ is assumed here, since a 
non-vanishing mean of the stochastic noise $\zeta^\alpha$ could just be included 
in the systematic Langevin forces 
$F^\alpha[S] = F^\alpha_{\rm rev}[S] + F^\alpha_{\rm rel}[S]$.
These incorporate reversible contributions that originate from the underlying
microscopic dynamics, i.e., Poisson brackets or commutators in a classical or
quantum-mechanical setting, and irreversible relaxational terms 
$F^\alpha_{\rm rel}[S] = - D_\alpha (i {\vec \nabla})^{a_\alpha} \, 
 \delta {\cal H}[S] / \delta S^\alpha$, where $a_\alpha = 0$ or $2$ respectively
for non-conserved and conserved stochastic fields.
Furthermore, the noise correlator $L^\alpha[S]$ may be an operator, as is the 
case for conserved variables, and could also be a functional of the slow fields
$S^\alpha$. 
In order for the dynamics to asymptotically reach the canonical thermal 
equilibrium distribution ${\cal P}_s[S]$ as $t \to \infty$, two fundamental 
conditions must be satisfied:
(i) For the relaxational terms, the set of Einstein relations 
$L^\alpha = k_{\rm B} T \, D_\alpha (i {\vec \nabla})^{a_\alpha}$ must hold, and
(ii) the probability current associated with the reversible Langevin forces should
be divergence-free in the space spanned by the hydrodynamic fields $S^\alpha$
\cite{Deker75}:
\begin{equation} 
  \int \! d^dx \, \sum_\alpha \frac{\delta}{\delta S^\alpha({\vec x})} \left( 
  F_{\rm rev}^\alpha[S] \ e^{- {\cal H}[S] / k_{\rm B} T} \right) = 0 \ . 
\label{divfre} 
\end{equation} 

The analysis of stochastic differential equations of the form \ref{langen}. with
noise \ref{noigen}. is conveniently pursued through a path integral representation
\cite{DeDominicis76, Janssen76, Bausch76, Janssen79, Folk06, Tauber14}.
The crucial assumption is that the noise history is a stochastic process with a
Gaussian distribution 
\begin{equation}
  {\cal W}[\zeta] \sim \exp \, \biggl[ - \frac14 \int \! d^dx \int_0^{t_f} \! dt 
  \, \sum_\alpha \zeta^\alpha({\vec x},t) \, \bigl( L^\alpha[S]({\vec x},t)]
  \bigr)^{-1} \zeta^\alpha({\vec x},t) \biggr] \ .
\label{noidis}
\end{equation}
Upon eliminating the noise $\zeta^\alpha$ via the Langevin equation \ref{langen}.
and further linearization by means of a Hubbard--Stratonovich transformation one
arrives at the probability distribution for the mesoscopic stochastic fields
${\cal P}[S] \sim \int \! {\cal D}[i {\widetilde S}] \, 
 e^{- {\cal A}[{\widetilde S},S]}$, with a statistical weight given by the 
Janssen--De~Dominicis response functional
\begin{equation}  
  {\cal A}[{\widetilde S},S] = \int \! d^dx \int_0^{t_f} \!\! dt \,
  \sum_\alpha {\widetilde S}^\alpha({\vec x},t) \left[ 
  \frac{\partial S^\alpha({\vec x},t)}{\partial t} - F^\alpha[S]({\vec x},t) 
  - L^\alpha[S]({\vec x},t) \, {\widetilde S}^\alpha({\vec x},t) \right] .
\label{janded}
\end{equation}
This resulting effective field theory contains two independent variables 
${\widetilde S}^\alpha$ and $S^\alpha$ in $d + 1$ dimensions, where the time-like 
direction of course plays a special role, since causality must be properly 
implemented.
For example, two-point correlation and dynamical response functions are given by
$C^{\alpha \beta}({\vec x},t;{\vec x\,}',t') = \left\langle S^\alpha({\vec x},t) 
 \, S^\beta({\vec x\,}',t') \right\rangle$ and 
$\chi^{\alpha \beta}({\vec x},t;{\vec x\,}',t') = \delta \left\langle 
 S^\alpha({\vec x},t) \right\rangle / \delta h^\beta({\vec x\,}',t')_{h = 0} 
 = D_\beta \big\langle S^\alpha({\vec x},t) \, (i {\vec \nabla})^{a_\beta} \,
 {\widetilde S}^\beta({\vec x\,}',t') \big\rangle$ for mere relaxational 
kinetics with $F^\alpha_{\rm rev}[S] = 0$.
A similar doubling of dynamical degrees of freedom occurs in the 
Keldysh--Baym--Kadanoff field theory formalism for non-equilibrium quantum
systems \cite{Kamenev11, Sieberer16}.

In isotropic magnetic systems, rotational invariance and the ensuing form of the
reversible Langevin forces fully determine the dynamical critical exponent $z$
\cite{Ma76, Hohenberg77, Folk06, Tauber14}.
For the critical dynamics of isotropic ferromagnets with conserved magnetization
(model J), one obtains $z = (d + 2 - \eta) / 2$ in dimensions $d \leq d_c' = 6$.
On the other hand, in planar ferromagnets (model E), isotropic antiferromagnets 
(model G), and superfluids (model F), a non-conserved order parameter couples
reversibly to a conserved mode; under the strong dynamic scaling scaling 
assumption that the characteristic relaxation times for all slow fields are 
governed by the same exponent, one finds $z = d / 2$ for $d \leq d_c = 4$.
Strong dynamic scaling also applies to the scalar model C, where a non-conserved
order parameter $S$ interacts with the conserved energy density $\rho$, and the 
dynamic exponent therefore can be expressed in terms of static critical exponents:
$z = 2 + \alpha / \nu \geq 2$, as $\alpha \geq 0$.
However, if one considers a $O(n)$-symmetric situation with $n \geq 2$ vector 
order parameter components, in fact $\alpha < 0$ and the energy density 
dynamically decouples in the critical regime: $z_\rho = 2$, whereas 
$z_S = 2 + c \, \eta$ as for model A \cite{Folk03, Folk04}.
For the analogous model D with a conserved order parameter, one always observes
weak dynamic scaling with $z_S = 4 - \eta$ as for model B, while
$z_\rho = {\rm min}\left( 2 , 2 + \alpha / \nu \right)$.
The critical dynamics of binary fluids or at the liquid-gas transition (model H) 
with a conserved scalar order parameter $S$ coupling to a conserved current 
${\vec j}$ constitutes another prominent example for weak dynamic scaling with 
$z_S > z_j$ satisfying the scaling relation $z_S + z_j = d + 2$ (for 
$d \leq d_c = 4$; $z_S = 4$ and $z_j = 2$ for $d > 4$).

\section{NON-EQUILIBRIUM CRITICAL DYNAMICS}

This section concerns the emerging dynamical scaling properties in near-critical 
systems that are forced out of equilibrium by either violating the 
detailed-balance conditions, or by initializing them in a far-from-equilibrium 
state and subsequently allowing them to relax.

\subsection{Critical Dynamics in the Absence of Detailed Balance}

Explicitly violating Einstein's relation between the relaxation rate and noise
strength, or breaking the divergence-free condition \ref{divfre}. will drive a 
dynamical system out of thermal equilibrium.
We are here concerned with the ensuing universal scaling features near a 
continuous phase transition that separates distinct non-equilibrium stationary 
states at asymptotically long times.
Note that in order to properly define such non-equilibrium phase transitions the
limit $t \to \infty$ must precede the tuning of external control parameters 
through the critical point, akin and in addition to the standard equilibrium 
requirement that the thermodynamic limit (infinite system size) must be taken 
first as well.

Considering first purely relaxational or model A kinetics, we see that an
arbitrary ratio $\Gamma / D = k_{\rm B} T_{\rm eff}$ in \ref{relmda}. and 
\ref{stnmda}. may serve to define an effective fluctuation temperature 
$T_{\rm eff}$.
The factor $T / T_{\rm eff}$ can then be absorbed into rescaled fields $S$ and
Landau--Ginzburg control parameters $h$, $r$, and $u$ in the Hamiltonian 
\ref{lgwham}.
At any ensuing critical point, $h$ and the renormalized counterpart $\tau$ of $r$
need to be set to zero, while the non-linear coupling $u$ approaches a universal
fixed-point value $u^*$.
The above rescaling hence only modifies the starting point for the renormalization
group flow, but leaves the asymptotic critical scaling behavior identical to that
of the equilibrium model A wherein detailed balance is manifestly encoded through
Einstein's relation.
Indeed, model A relaxational kinetics turns out quite robust against 
non-equilibrium perturbations \cite{Haake84, Grinstein85}; this is even true when 
these explicitly break the $Z_2$ symmetry for the Ising model with Glauber spin
dynamics \cite{Bassler94}.
Consequently, model A critical dynamics emerges as one of the prominent and
ubiquitous universality classes that describe various non-equilibrium phase
transitions.

Remarkably, this statement extends to the critical scaling features of the complex
Ginzburg--Landau equation for a complex order parameter field $\psi$ with additive
white noise:
\begin{eqnarray} 
  &&\frac{\partial \psi({\vec x},t)}{\partial t} = - D \left[ \tau + i \tau' - 
  \left( 1 + i r_k \right) {\vec \nabla}^2 + \frac{u}{6} \left( 1 + i r_u \right) 
  |\psi({\vec x},t)|^2 \right] \psi({\vec x},t) + \zeta({\vec x},t) \ , \nonumber 
  \\ &&\big\langle \zeta({\vec x},t) \big\rangle = 0 
  = \left\langle \zeta({\vec x},t) \, \zeta({\vec x\,}',t') \right\rangle \, , 
  \quad \left \langle \zeta^*({\vec x},t) \, \zeta({\vec x\,}',t') \right\rangle 
  = 4 \Gamma \, \delta({\vec x} - {\vec x\,}') \, \delta(t - t') \ . \quad
\label{cogleq}
\end{eqnarray}
This stochastic dynamics describes, e.g., the synchronization transition of 
coupled non-linear oscillators subject to random external drive \cite{Risler05}, 
and rather generically spontaneous structure formation out of equilibrium 
\cite{Cross93, Cross09}, which pertains to the population dynamics of cyclically 
competing species and evolutionary game theory \cite{Frey10}.
Eq.~\ref{cogleq}. also represents a noisy quantum-mechanical Gross--Pitaevskii 
equation that captures driven-dissipative Bose--Einstein condensation for 
interacting bosonic quasi-particles, e.g., exciton-polaritons in optically pumped 
semiconductor quantum wells or ultracold ions trapped in optical lattices 
\cite{Sieberer13, Sieberer14, Diehl14}.
At the (bi-)critical point where $\tau \sim \tau' \to 0$ vanish simultaneously, 
the general scaling form for the dynamical response and correlation function 
becomes
\begin{eqnarray} 
  &&\chi(\tau,{\vec q},\omega) = |{\vec q}\,|^{- 2 + \eta} 
  \left( 1 + i a \, |{\vec q}\,|^{\eta - \eta_c} \right)^{-1} 
  {\hat \chi}_\pm\!\left( {\vec q} \, \xi , \omega / \left[ D \, |{\vec q}\,|^z 
  (1 + i a \, |{\vec q}\,|^{\eta - \eta_c}) \right] \right) \, , \nonumber \\
  &&C(\tau,{\vec q},\omega) = |{\vec q}\,|^{- 2 - z + \eta'} {\hat C}_\pm\!\left( 
  {\vec q} \, \xi , a \, |{\vec q}\,|^{\eta - \eta_c} , \omega / D \, 
  |{\vec q}\,|^z \right) \, . 
\label{coglsc}
\end{eqnarray}
Since this system too thermalizes in the critical regime, the 
fluctuation-dissipation relation \ref{fldist}. is restored there, whence
$\eta = \eta'$ and the static as well as dynamical critical exponents $\eta$, 
$\nu$, and $z$ in \ref{coglsc}. are identical to those of the equilibrium 
two-component model A.
The ultimate disappearance of quantum coherence in \ref{cogleq} is captured 
through the universal correction-to-scaling exponent $\eta_c = c' \eta$, which has
been computed via the functional renormalization group \cite{Sieberer13, 
Sieberer14} and perturbatively as $c' = 1 - 4 \ln (4/3) + O(\epsilon) < 0$ 
\cite{Diehl14}. 

The above rescaling arguments also apply to both model B for diffusive relaxation
kinetics, and to model J for the critical dynamics of isotropic ferromagnets, as 
long as any detailed-balance violation occurs isotropically both in order 
parameter and real space \cite{Racz97, Santos99, Akkineni02}.
However, in such systems with conserved order parameter, one can implement 
non-equilibrium perturbations in an anisotropic manner.
For example, in model B, one may allow anisotropic diffusive relaxation via
replacing $- D \, {\vec \nabla}^2 \to - D_\parallel \, {\vec \nabla}_\parallel^2
 - D_\perp \, {\vec \nabla}_\perp^2$, and moreover set the conserved noise 
strength to $\Gamma = - {\widetilde D}_\parallel {\vec \nabla}_\parallel^2 
 - {\widetilde D}_\perp {\vec \nabla}_\perp^2$ with 
 ${\widetilde D}_\parallel / D_\parallel > {\widetilde D}_\perp / D_\perp$,
i.e., effective temperatures $T_\perp < T_\parallel$ in the two distinct spatial
sectors.
As the critical temperature $T_c$ is approached from above, consequently the 
transverse order parameter fluctuations soften first, whereas the longitudinal 
spatial sector remains non-critical.
In the critical region for this so-called two-temperature model B or model B with
anisotropic random drive, one may thus disregard the longitudinal noise (i.e., let
${\widetilde D}_\parallel \to 0$) and non-linear fluctuations 
\cite{Schmittmann91, Schmittmann93}.
The remaining terms in the resulting Langevin equation can be straightforwardly
recast in the form of an equivalent equilibrium model B
\begin{equation}
  \frac{\partial S({\vec x},t)}{\partial t} = D_\perp \, {\vec \nabla}_\perp^2 \,
  \frac{\delta {\cal H}_{\rm eff}[S]}{\delta S({\vec x},t)} + \zeta({\vec x},t)\ ,
\label{rdrmdb}
\end{equation}
albeit with a coarse-grained effective Hamiltonian that contains spatially 
long-range correlations akin to those in uniaxial dipolar magnets or ferroelastic 
materials \cite{Schwabl96}:
\begin{equation}
  \frac{{\cal H}_{\rm eff}[S]}{k_{\rm B} T} = \int \! \frac{d^dq}{(2 \pi)^d} \ 
  \frac{c \, {\vec q\,}_\parallel^2 + {\vec q\,}_\perp^2 (r + {\vec q\,}_\perp^2)}
  {2 \, {\vec q\,}_\perp^2} \, |S({\vec q})|^2 
  + \frac{\tilde u}{4!} \int \! d^dx \, S({\vec x})^4 \ ,
\label{ttmdbh}
\end{equation}
where ${\tilde u} = u \, {\widetilde D}_\perp / D_\perp$.
Thus anisotropic scaling ensues, e.g., for the dynamical susceptibility
\begin{equation}
  \chi(\tau, {\vec q}_\parallel, {\vec q}_\perp, \omega) = 
  |{\vec q}_\perp|^{- 2 + \eta} \, 
  {\hat \chi}_\pm\!\left( \tau / |{\vec q}_\perp|^{1 / \nu} , 
  \sqrt{c} \, {\vec q}_\parallel / |{\vec q}_\perp|^{1 + \Delta} , 
  \omega / D \, |{\vec q}_\perp|^z \right) \, ,
\label{ttmdbs}
\end{equation}
where an additional critical anisotropy exponent $\Delta$ has been introduced.
Since only the $d_\parallel$-dimensional transverse spatial sector softens as 
$\tau \to 0$, the upper critical dimension is lowered to $d_c = 4 - d_\parallel$,
as can be inferred from dimensional analysis ($d_\parallel + d_\perp = d$): 
With $[{\vec q}_\perp] = \mu$ and $[{\vec q}_\parallel] = \mu^2$, we have 
$[r] = \mu^2$, $[S({\vec x})] = \mu^{d_\parallel + (d_\perp - 2) / 2}$, and 
$[{\tilde u}] = \mu^{4 - 2 d_\parallel - d_\perp}$. 
The order parameter conservation law enforces the exact scaling relations 
$z = 4 - \eta$ as in equilibrium, yet with an altered static Fisher exponent 
$\eta$, and $\Delta = 1 - \eta / 2 = - 1 + z / 2$.

Investigating the effect of detailed-balance violations on the near-equilibrium
critical dynamics universality classes yields that models with non-conserved order
parameter are generally quite stable against such non-equilibrium perturbations,
whereas spatially anisotropic order parameter noise correlations may induce 
drastic deviations from equilibrium behavior \cite{Santos99, Santos02, Akkineni02,
Akkineni04, Tauber14}; this is true especially in the presence of reversible 
couplings to other conserved modes, for which the equilibrium condition 
\ref{divfre}. becomes invalidated.

\subsection{Non-Equilibrium Critical Relaxation and Aging Scaling}

Quite generally, when a stochastic dynamical system is prepared in a starting
configuration that differs considerably from its asymptotic long-time stationary 
state, it retains memory of the initial conditions in a transient time window that
typically extends to the scale of its characteristic relaxation time $t_c$.
For times $t_m \ll t' < t < t_c$, where $t_m$ denotes any microscopic time scale, 
the system thus cannot have reached stationarity, and two-time observables
will depend on both waiting and observation times $t'$ and $t$ \cite{Henkel10}.
In glassy systems with exceedingly slow relaxation, the resulting physical-aging
features become prominent and afford a very useful means to probe intrinsic
dynamical processes and their correlations.
Moreover, when exponential temporal decay is effectively replaced by algebraic 
power laws, $t_c \to \infty$, and the aging time window dominates the system's
entire relaxation kinetics.
A classical example is non-equilibrium phase ordering, where a system may be 
prepared in an initially fully disordered state (corresponding to a large 
temperature $T \gg T_c$), but is then suddenly quenched into the ordered phase
($T < T_c$).
It then quickly forms domains wherein the order parameter locally acquires one of
the allowed degenerate values, which subsequently grow and merge through a slow 
dynamical coarsening process \cite{Bray94}.
The characteristic domain size grows with time according to a power law 
$L(t) \sim \left( D t \right)^{1 / z}$ with a dynamical scaling exponent $z$.
For example, $z = 2$ for non-conserved model A relaxation, whereas for a 
conserved scalar order parameter with diffusive kinetics $z = 3$.
In contrast, for the $O(n)$-symmetric vector model B one obtains $z = 4$ for
$n \geq 3$, with logarithmic corrections for the borderline two-component case: 
$L(t) \sim \left[ D t \ln (D t) \right]^{1/4}$ \cite{Bray94, Tauber14}.
 
In the vicinity of a critical point, the correlation length $\xi$ replaces the
characteristic domain size $L$ as the ultimately diverging length scale, and as
the relaxation time $t_c(\tau) \sim \xi(\tau)^z \sim |\tau|^{- z \nu}$ also
diverges upon approaching the continuous phase transition, critical slowing-down
drastically widens the physical aging time window.
By means of a dynamic renormalization group analysis in conjunction with a 
short-time operator product expansion, one can derive the following simple-aging
scaling laws for the two-time dynamical response and correlation functions
\cite{Janssen89, Janssen92, Calabrese05, Henkel10, Tauber14}:
\begin{eqnarray}
  &&\chi(\tau,{\vec q}, t' \ll t, t) = |{\vec q}\,|^{z - 2 + \eta} \left( t / t'
  \right)^\theta {\hat \chi}_0\!\left( {\vec q} \, \xi , |{\vec q}\,|^z t \right) 
  \, , \nonumber \\
  &&C(\tau,{\vec q}, t' \ll t, t) = |{\vec q}\,|^{- 2 + \eta} \left( t / t' 
  \right)^{\theta - 1} {\hat C}_0\!\left( {\vec q} \, \xi, |{\vec q}\,|^z t 
  \right) \, , 
\label{cragsr} 
\end{eqnarray}
where the fluctuation-dissipation theorem \ref{fldist}. was invoked.
The time evolution of the order parameter displays the intriguing initial-slip 
scaling behavior
\begin{equation}
  \phi(t) = \phi_0 \, t^{\theta'} 
  {\hat S}_0\bigl( \phi_0 \, t^{\theta' + \beta / z \nu} \bigr) \ . 
\label{inslip}
\end{equation}

For model A, both $\theta$ and $\theta' = \theta - 1 + (2 - \eta) / z > 0$, whence
the critical order parameter grows initially, and only at late times decays to 
zero according to $\phi(t \to \infty) \sim t^{- \beta / z \, \nu}$.
Similarly, the dynamic susceptibility decays asymptotically as
$\chi\!\left( \tau, {\vec x}, t' = 0, t \to \infty \right) \sim 
 t^{- \lambda_R / z}$ with $\lambda_R = d - z \, \theta'$.
The critical initial-slip exponent $\theta'$ here represents a genuinely 
independent scaling exponent that is related to singular behavior specific to the
initial time sheet at $t' = 0$ \cite{Janssen89, Janssen92}.
For a conserved order parameter, however, no new singularities can appear as 
$t' \to 0$, and correspondingly $\theta$, $\theta'$, $\lambda_R$ may be expressed
in terms of the other critical exponents:
For model B, one simply finds $\theta = \theta' = 0$ and $\lambda_R = d + 2$
\cite{Oerding93}; on the other hand for, e.g., model J capturing the critical 
dynamics of isotropic ferromagnets, 
$\theta = 1 - (4 - \eta) / z = (d - 6 + \eta) / (d + 2 - \eta)$ \cite{Janssen93}.
In these situations, the critical exponents that describe the ultimate stationary
scaling behavior can already be efficiently accessed in numerical simulations 
through a careful study of the much earlier aging scaling of two-time observables
\ref{cragsr}., or via the related order parameter initial-slip behavior 
\ref{inslip}. \cite{Zheng98}.

\section{SCALE INVARIANCE AND PHASE TRANSITIONS IN DRIVEN SYSTEMS}

Let us now turn our attention to the emergence of generic scale invariance in
driven stationary states far from thermal equilibrium, and to the dynamic scaling
properties near genuine non-equilibrium continuous phase transitions.

\subsection{Driven Diffusive Systems}

We first address the intriguing scale-invariant features of driven lattice gases 
\cite{Schmittmann95, Marro99} with hard-core repulsive interactions, which in one 
dimension are referred to as asymmetric exclusion processes \cite{Derrida98,
Stinchcombe01}:
Particles move via nearest-neighbor hopping that is biased along a specified
drive ($\parallel$) direction, subject to an exclusion constraint; i.e., only at 
most a single particle is permitted on each site.
The allowed occupation numbers $n_i = 0, 1$ can naturally be mapped onto binary or
Ising spin variables $\sigma_i = 2 n_i - 1 = \mp 1$.
Here, we only consider driven systems with periodic boundary conditions, for which
the biased diffusive propagation generates a non-zero mean particle current. 
At long times, the kinetics thus reaches a non-equilibrium steady state which is 
in fact governed by algebraic rather than exponential temporal correlations. 
The stationary non-equilibrium dynamics thus displays generic scale invariance, 
without the need of tuning the system to a special critical point.
Extensions of these simple, but phenomenologically rich systems serve as 
paradigmatic models for a wide variety of directed stochastic transport problems 
in biology and biochemistry \cite{Chou11}. 

To construct a coarse-grained description for the non-equilibrium steady state of 
this system of particles with conserved density $\rho({\vec x},t)$ and hard-core 
repulsion, driven along the $\parallel$ direction on a $d$-dimensional lattice 
\cite{Janssen86, Schmittmann95}, we start with the continuity equation 
\begin{equation}
  \frac{\partial S({\vec x},t)}{\partial t} 
  + {\vec \nabla} \cdot {\vec J}({\vec x},t) = 0 \ ,
\label{ddscon}
\end{equation}
where the scalar field $S({\vec x},t) = 2 \rho({\vec x},t) - 1$ represents a local
magnetization in the spin language, whose mean remains fixed at 
$\langle S({\vec x},t) \rangle = 0$ for a half-filled lattice.
To specify the current density ${\vec J}({\vec x},t)$, we assume a mere noisy 
diffusion current in the $d_\perp$-dimensional transverse sector 
($d_\perp = d - 1$).
Along the drive, however, both bias and exclusion are crucial: $J_\parallel = - c 
 \, D \, \nabla_\parallel S + 2 D \, g \, \rho \left( 1 - \rho \right) + \zeta$,
where $c$ measures the ratio of diffusivities parallel and transverse to the net 
current.
In the comoving reference frame with 
$\langle J_\parallel({\vec x},t) \rangle = 0$, therefore
\begin{eqnarray}
  &&{\vec J}_\perp({\vec x},t) = - D \, {\vec \nabla}_\perp S({\vec x},t) 
  + {\vec \eta}({\vec x},t) \ , \nonumber \\
  &&J_\parallel({\vec x},t) = - c \, D \, \nabla_\parallel S({\vec x},t) 
  - \frac{D \, g}{2} \, S({\vec x},t)^2 + \zeta({\vec x},t) \ , 
\label{ddscur}   
\end{eqnarray}
with $\langle \eta_i({\vec x},t) \rangle = 0 = \langle \zeta({\vec x},t) \rangle$,
and the noise correlations
\begin{eqnarray} 
  &&\left\langle \eta_i({\vec x},t) \, \eta_j({\vec x}\,',t') \right\rangle 
  = 2 D \, \delta({\vec x} - {\vec x}\,') \, \delta(t - t') \, \delta_{ij} \ , 
  \nonumber \\ 
   &&\left\langle \zeta({\vec x},t) \, \zeta({\vec x}\,',t') \right\rangle 
  = 2 D \, {\tilde c} \, \delta({\vec x} - {\vec x}\,') \, \delta(t - t') \ . 
\label{ddsnoi} 
\end{eqnarray}

Einstein's relations which link the noise strengths and the relaxation rates need 
of course not be satisfied in the ensuing non-equilibrium steady state.
Yet for the transverse sector, say, one may formally enforce such a connection
through a straightforward rescaling of the field $S$.
The deviation from the Einstein relation in the drive direction is then encoded in
\ref{ddscur}. and \ref{ddsnoi}. through the ratio $0 < w = {\tilde c}/c$.
The resulting Langevin equation is akin to the critical linear model B with 
anisotropic diffusion and noise, but with a non-linear drive term that breaks both
the system's spatial inversion and Ising $Z_2$ symmetries.
The driven diffusive dynamics hence is generically scale invariant, with the 
dynamic response and correlation functions satisfying anisotropic scaling laws as 
in \ref{ttmdbs}. at criticality ($\tau = 0$).
Dimensional analysis with $[q_\parallel] = [{\vec q}_\perp] = \mu$, 
$[\omega] = [t]^{-1} = \mu^2$, $[D] = [c] = [{\tilde c}] = \mu^0$, and 
$[S({\vec x},t)] = \mu^{d / 2} $ yields $[g] = \mu^{1 - d / 2}$, indicating 
$d_c = 2$ as the upper critical dimension.
Yet since the non-linear term only affects the fluctuations in the direction along
the drive, the transverse sector is characterized by Gaussian scaling 
exponents $\eta = 0$ and $z = 2$ \cite{Janssen86, Schmittmann95}.

The mesoscopic stochastic differential equation \ref{ddscon}. with \ref{ddscur}. 
displays an emergent symmetry that is not explicit in the underlying microscopic 
lattice model, namely it remains invariant under the generalized Galilean 
transformation
\begin{equation}
  S\!\left( {\vec x}_\perp, x_\parallel, t \right) \to 
  S'\!\left( {\vec x\,}_\perp', x_\parallel', t' \right) 
  = S\!\left( {\vec x}_\perp, x_\parallel - D \, g \, v \, t, t \right) - v \ .
\label{galtra}
\end{equation}  
This in fact fixes the anisotropy exponent exactly to $\Delta = (2 - d) / 3$ for 
$d \leq d_c = 2$, and hence the longitudinal dynamic scaling exponent 
$z_\parallel = z / (1 + \Delta) = 6 / (5 - d)$ \cite{Janssen86, Schmittmann95, 
Tauber14}.
For the asymmetric exclusion process in one dimension, this yields 
$z_\parallel = 3/2$.
Moreover, the renormalization group analysis demonstrates that at the fixed point 
the Einstein ratio assumes the equilibrium value $w^* = 1$ \cite{Janssen86, 
Tauber14}.
For $d = 1$, the Langevin dynamics \ref{ddscur}., \ref{ddsnoi}. then maps to the
noisy Burgers equation for randomly stirred fluids \cite{Forster77} for the 
stochastic velocity field $u(x,t) = - S(x,t)$.
Indeed, a straightforward calculation demonstrates that the canonical probability 
distribution ${\cal P}_s[u] \sim \exp \left[- \frac12 \int u(x)^2 dx \right]$
with the fluid's kinetic energy satisfies the potential condition \ref{divfre}.
\cite{Janssen99}.
The anomalous dynamic scaling for driven diffusive systems for $d \leq d_c = 2$ 
can also be inferred from their non-stationary relaxation.
As for model B, the fundamental particle conservation law implies that the ensuing
simple aging kinetics for the dynamic correlation function \ref{cragsr}. does not 
require a new scaling exponent, but $\theta = - \Delta$ and 
$\lambda_C / z = (d + \Delta) / 2 = (d + 1) / 3$ \cite{Daquila11, Tauber14}.

Even richer scaling behavior ensues if in addition to the hard-core repulsion, 
nearest-neighbor attractive interactions are incorporated to the driven Ising 
lattice gas \cite{Katz83, Katz84, Schmittmann95, Marro99}.
This Katz--Lebowitz--Spohn model displays a genuine non-equilibrium continuous 
phase transition in dimensions $d \geq 2$, from a disordered phase, governed by 
the scaling laws described above, to an ordered state characterized by phase 
separation into low- and high-density regions, with the phase boundary oriented 
along the drive and resulting net particle current direciton.
As the hopping bias vanishes ($g = 0$), this continuous phase transition is of 
course described by the $d$-dimensional ferromagnetic equilibrium Ising model.
In the continuum description, we essentially need to add the drive non-linearity
from \ref{ddscur}. to the model B Langevin equation for a conserved order 
parameter \cite{Janssen87, Leung86, Schmittmann95, Tauber14}.
As in the two-temperature model B above, we must only retain non-linear 
fluctuations in the transverse spatial sector, whence we arrive at the 
coarse-grained stochastic differential equation
\begin{equation} 
  \frac{\partial S({\vec x},t)}{\partial t} = D \left[ c \nabla_\parallel^2 
  + {\vec \nabla}_\perp^2 \left( r - {\vec \nabla}_\perp^2 \right) \right] 
  S({\vec x},t) + \frac{D \, g}{2} \, \nabla_\parallel S({\vec x},t)^2 
  + \frac{D \, {\tilde u}}{6} \, {\vec \nabla}_\perp^2 S({\vec x},t)^3 
  + \zeta({\vec x},t) \ ,
\label{klslan}
\end{equation} 
with conserved noise that satisfies $\langle \zeta({\vec x},t) \rangle = 0$ and
\begin{equation}
  \left\langle \zeta({\vec x},t) \, \zeta({\vec x\,}',t') \right\rangle = - 2 D \,
  {\vec \nabla}_\perp^2 \, \delta({\vec x} - {\vec x\,}') \, \delta(t - t') \ .
\label{klsnoi}
\end{equation}
The critical Katz--Lebowitz--Spohn model thus contains non-vanishing three-point 
correlations, which are absent in the high-temperature phase of the randomly 
driven model B.

\begin{figure}[h]
\includegraphics[width=8.5cm,clip]{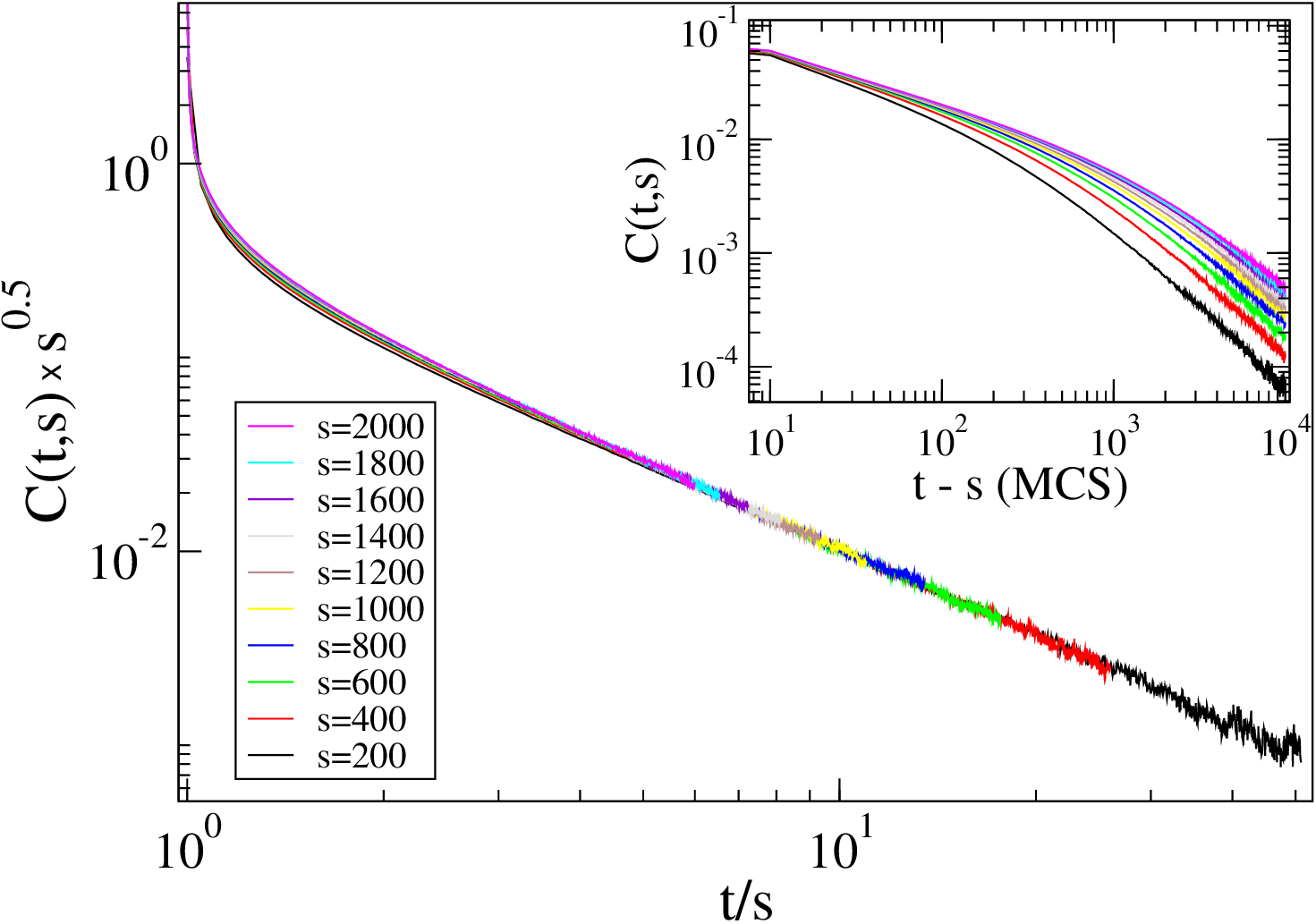} \vskip -0.2cm
\caption{Critical aging scaling in Monte Carlo simulation data for the two-time 
   density auto-correlation function in the Katz--Lebowitz--Spohn model on a
   $125,000 \times 50$ rectangular lattice following a quench from a 
   high-temperature disordered configuration. 
   The inset shows the same data plotted as function of the time difference $t-s$, 
   demonstrating broken time translation invariance (Figure reproduced with 
   permission from: G.L.~Daquila, 2011 Ph.D. dissertation, Virginia Tech).}
\label{fig1}
\end{figure}
Moreover, the associated upper critical dimensions of these models differ; for 
dimensional analysis with $[q_\parallel] = [{\vec q}_\perp]^2 = \mu^2$ and 
$[c] = \mu^0$ gives for the driven Ising lattice gas
$[S({\vec x},t)] = \mu^{- 1 + d / 2}$ and consequently 
$[{\tilde u}] = \mu^{3 - d}$, $[g] = \mu^{(5 - d) / 2}$.
Therefore $d_c = 5$, and the static non-linearity ${\tilde u}$ is irrelevant as 
compared with the drive $g$.
While it may be omitted for the determination of the asymptotic universal scaling 
laws, at least in scaling functions one may not simply set ${\tilde u} = 0$, since
this dangerously irrelevant coupling is of course responsible for the occurrence 
of the phase transition.
As in the non-critical driven lattice gas, the transverse sector is not affected
by the drive non-linearity, and hence characterized by Gaussian model B critical 
exponents $\eta = 0$, $\nu = 1 / 2$, and $z = 4$ in \ref{ttmdbs}.
In addition, Galilean invariance \ref{galtra}. holds and for $d \leq d_c = 5$
implies the exact anisotropy exponent $\Delta = (8 - d) / 3$, and therefrom 
$\nu_\parallel = \nu \, (1 + \Delta) = (11 - d) / 6$ and 
$z_\parallel = z / (1 + \Delta) = 12 / (11 - d)$ \cite{Janssen87, Leung86, 
Schmittmann95, Tauber14}.
Numerically, even on properly constructed anisotropic simulation domains, 
exceedingly long crossover times leading towards the asymptotic stationary scaling
regime prohibit accurate determination of the critical exponents in driven Ising 
lattice cases \cite{Caracciolo04}.
Yet satisfactory dynamic aging scaling collapse can be achieved for the two-time 
density autocorrelation function with 
$\lambda_C = d - 2 + \eta + \Delta = 2 \, (d + 1) / 3$, as demonstrated in the
Monte Carlo simulation data in Fig.~\ref{fig1} \cite{Daquila12, Tauber14}.

\subsection{Driven Interfaces: Kardar--Parisi--Zhang Model and Variants}

Generic scale invariance emerges remarkably frequently in out-of-equilibrium
systems.
Aside from driven lattice gases, prominent examples include moving interfaces,
pulled or pushed through materials by external driving, as well as surface growth
under non-equilibrium conditions \cite{Krug92, Barabasi95, Halpin95, Krug97}.
For isotropic materials or substrates, and in the absence of long-range 
correlations in the thermal noise or random particle deposition processes, the
scaling behavior of these systems is generically described by the 
Kardar--Parisi--Zhang model \cite{Kardar86}, captured by the non-linear Langevin
equation
\begin{eqnarray}
  &&\frac{\partial S({\vec x},t)}{\partial t} = D \, {\vec \nabla}^2 S({\vec x},t)
  + \frac{D \, g}{2} \left[ {\vec \nabla} S({\vec x},t) \right]^2 
  + \zeta({\vec x},t) \ , 
\label{kpzlan} \\
  &&\big\langle \zeta({\vec x},t) \big\rangle = 0 \ , \quad
  \left\langle \zeta({\vec x},t) \, \zeta({\vec x\,}',t') \right\rangle 
  = 2 D \, \delta({\vec x} - {\vec x\,}') \, \delta(t - t') \ .
\label{kpznoi}
\end{eqnarray}
The scalar field $S({\vec x},t)$ represents the interface or surface height
fluctuations relative to its mean position that moves or grows linear with time
$t$. 
The $d$-dimensional substrate is parametrized by the coordinates ${\vec x}$, and 
a unique height profile function is surmised, i.e., any overhangs are neglected 
(or adequately coarse-grained). 
The non-linear term $\sim g$ describes curvature-driven propagation or growth.
The height fluctuations are then scale-invariant at sufficiently large length 
scales up to $|{\vec x}| \leq L(t) \sim (D \, t)^{1/z}$ with dynamic exponent $z$.
The dynamical height correlation function should thus obey the critical 
($\tau = 0$) scaling law \ref{dynscr}., which in this context is usually written 
in terms of a roughness exponent $\chi$ \cite{Family85}:
\begin{equation}
  C({\vec x},t) = |{\vec x}|^{2 \chi} \, {\hat C}(D \, t / |{\vec x}|^z) \ , 
  \quad \chi = \frac12 \left(2 - d - \eta \right) \, .
\label{intsca}
\end{equation}

The associated linear growth model ($g = 0$) or Edwards--Wilkinson equation 
\cite{Edwards82} is just a noisy diffusion equation or Gaussian model A at 
criticality.
Its effectively equilibrium kinetics tends towards the Gaussian stationary
probability distribution
\begin{equation}
  {\cal P}_s[S] \sim \exp \left( - \frac12 \int \left[ {\vec \nabla} S({\vec x}) 
  \right]^2 d^dx \right) \, ,
\label{ewstat}  
\end{equation} 
whence the corresponding scaling exponents are $\eta = 0$ and $z = 2$.
The roughness exponent is therefore $\chi = 1/2$ in one dimension, while the
interface becomes flat ($\chi = 0$) for $d \geq 2$. 
Indeed, $d_c = 2$ represents the critical dimension for this problem, as can be
inferred from direct scaling analysis: $[S({\vec x},t)] = \mu^{(d - 2) / 2}$, and
$[g] = \mu^{1 - d / 2}$. 
Moreover, the substitution ${\vec u}({\vec x},t) = - {\vec \nabla} S({\vec x},t)$
transforms the Kardar--Parisi--Zhang equation to the $d$-dimensional noisy 
Burgers equation for a vorticity-free velocity field, 
${\vec \nabla} \times {\vec u}({\vec x},t) = 0$.
The fluid dynamics invariance with respect to Galilean transformations, 
c.f.~\ref{galtra}., maps onto (infinitesimal) tilt symmetry for the interface
problem \cite{Medina89}: $S({\vec x},t) \to S'({\vec x\,}',t') = 
 S({\vec x} - D \, g \, {\vec v} \, t, t) - {\vec v} \cdot {\vec x}$.
Since the height field scales with the roughness exponent $\chi$, demanding this 
invariance to hold under scale transformations enforces the scaling relation 
$\chi + z = 2$.

\begin{figure}[h] \vskip -0.2cm
\includegraphics[width=8.5cm]{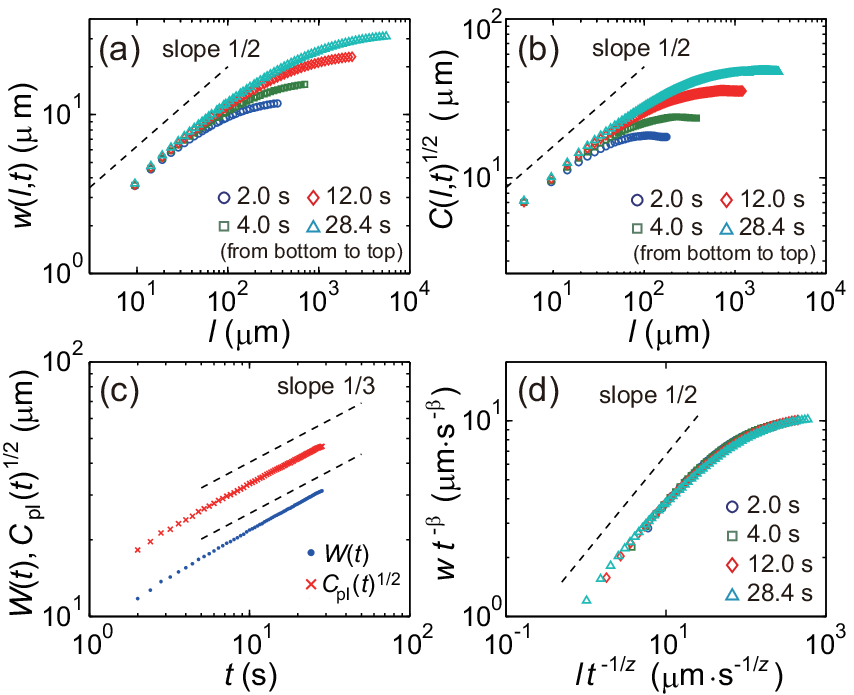} \vskip -0.2cm
\caption{Experimental confirmation of Kardar--Parisi--Zhang dynamical scaling in 
   electroconvective front propagation in turbulent nematic liquid crystals:
   (a) interface width $w(l,t)$ and (b) height difference correlation function 
   $C(l,t)$ as function of length scale $l = |\vec x|$ for various times $t$ 
   indicating initial growth with the roughness exponent $\chi = 1/2$;
   (c) time evolution of the overall width $W(t)$ and the plateau level 
   $C_{\rm pl}(t)$; and (d) Family--Vicsek scaling \cite{Family85} collapse 
   $w(l,t) = t^{\beta} {\hat w}(l \, t^{-1/z})$ with $\beta = \chi / z = 1/3$.
   (Figures reproduced with permission from Ref.~\cite{Takeuchi10}, 
   DOI: 10.1103/PhysRevLett.104.230601; 
   copyright 2010 by the American Physical Society).}
\label{fig2}
\end{figure}

Specifically in one dimension, the stationary distribution \ref{ewstat}. pertains 
even to the non-linear Langevin equation \ref{kpzlan}., with the Hamiltonian just 
representing the Burgers fluid's kinetic energy \cite{Medina89}.
The ensuing exact exponent values $\eta = 0$ and $\chi = 1/2$ imply $z = 3/2$, 
which of course coincides with the dynamical exponent $z_\parallel$ for the 
driven lattice gas.
An explicit dynamic renormalization group analysis confirms these results 
\cite{Frey94, Frey96, Tauber14}. 
It also allows an investigation of the non-equilibrium relaxation kinetics 
starting from an initially flat interface; the universal aging properties are 
again fully set by the stationary scaling exponents \cite{Krech97, Daquila11, 
Henkel12}.
There are convincing experimental realizations for the Kardar--Parisi--Zhang
scaling in $1 + 1$ dimensions that range from non-equilibrium surface growth, 
e.g., in electrodeposition \cite{Schilardi99}, to flame front propagation in slow 
paper combustion \cite{Myllys01}, and turbulent dynamics in the electroconvection 
of nematic liquid crystals \cite{Takeuchi10, Takeuchi12}; data and scaling plots
for the latter are depicted in Fig.~\ref{fig2}.

In dimensions $d \geq d_c = 2$, the Kardar--Parisi--Zhang equation displays even 
richer behavior, namely a non-equilibrium roughening transition that separates a 
smooth interface governed by the linear Edwards--Wilkinson equation from a rough 
phase, for which the non-linear coupling $g$ diverges in a perturbative 
renormalization group analysis \cite{Medina89, Frey94, Lassig95, Lassig98, 
Wiese98, Janssen99}.
The scaling properties of this strong-coupling rough phase can however be 
successfully accessed through a non-perturbative numerical renormalization group 
approach \cite{Canet10, Canet11, Kloss12}.
Note that $d_c = 2$ thus plays the role of a lower critical dimension for the 
existence of the roughening phase transition.
Intriguingly, the Cole--Hopf transformation 
$S({\vec x},t) = k_{\rm B} T \, \ln Z({\vec x},t) / D \, g \, \varepsilon$ 
\cite{Lassig95, Lassig98, Wiese98, Janssen99} maps the stochastic differential 
equation \ref{kpzlan}. with additive white noise \ref{kpznoi}. to an 
imaginary-time Schr\"odinger equation for the canonical partition function 
$Z({\vec x},t)$ of a directed polymer, described by its trajectory ${\vec x}(t)$ 
along the $t$ direction, and with elastic line tension $\varepsilon$ at 
temperature $T$ that is subject to a Gaussian-distributed, spatially uncorrelated
random pinning potential \cite{Kardar87, Fisher91, Hwa94}.
A simple scaling argument that demands that the elastic and pinning energies 
should both scale marginally at the roughening transition gives the critical 
exponent values $\chi_c = 0$ and $z_c = 2$ \cite{Doty92}.
This can be confirmed in a computation to all orders in a dimensional 
$d = 2 + \epsilon$ expansion, along with the correlation length exponent
$\nu_c = 1 / (d - 2)$ \cite{Lassig95, Wiese98, Janssen99, Tauber14}.

These considerations form the basis for more detailed investigations of the 
collective statistical mechanics and dynamics of interacting directed polymers in
random media \cite{Kardar98}, with technologically relevant applications to the 
vortex and Bragg glass phases in type-II superconductors with point pinning 
centers \cite{Nattermann00}.
Other broad extension include the non-equilibrium dynamics of extended elastic 
manifolds driven through random media \cite{Balents98, Fisher98} that display  
continuous depinning transitions, whose theoretical analysis requires functional
renormalization group tools \cite{Doussal04}, featuring scale-invariant avalanche
kinetics \cite{Doussal13}.

It turns out that anisotropies in the non-linear growth and relaxation terms, 
which would naturally be expected to occur in real surfaces, constitute relevant 
perturbations for the Kardar--Parisi--Zhang equation in dimensions $d > 2$, and 
lead to rich phenomena that include the possibility of first-order roughening 
transitions and multi-critical behavior \cite{Tauber02}.
The anisotropic Kardar--Parisi--Zhang equation also describes driven-dissipative
Bose--Einstein condensation in two dimensions \cite{Altman15}.
In many experiments on growing surfaces under non-equilibrium conditions, surface
diffusion plays a crucial role to relax spatial inhomogeneities.
Akin to model B for diffusive relaxational critical dynamics, so-called conserved
Kardar--Parisi--Zhang model variants thus incorporate an additional Laplacian 
$- {\vec \nabla}^2$ in front of the systematic contributions to the Langevin 
equation \ref{kpzlan}.; since the noise correlations are not constrained by a 
fluctuation-dissipation relation, one may then either consider the 
Sun--Gao--Grant model with conserved ($a = 2$) noise \cite{Sun89}, or the 
Wolf--Villain model with non-conserved ($a = 0$) shot noise \cite{Wolf90}.
In either case, the dynamic scaling exponent becomes $z = 4 - \eta$, and is 
related to the roughness exponent via the scaling relation
$\chi = (z - a - d) / 2 = (4 - a - d - \eta) / 2$ \cite{Hannes97, Tauber14}.

\section{REACTING PARTICLE SYSTEMS: SCALING AND PHASE TRANSITIONS}

This final section describes scale-invariant correlation-dominated relaxation 
kinetics in reacting particle systems \cite{Kuzovkov88, Ovchinnikov89, 
Krapivsky10}, and discusses active-to-absorbing state transitions in 
reaction-diffusion and simple epidemic or population dynamics models 
\cite{Hinrichsen00, Odor04, Henkel08, Tauber14}.

\subsection{Scale Invariance in Diffusion-Limited Reactions}

In chemical reactions, individual particles of species $A, B, \ldots$ are 
annihilated, created, or transformed either spontaneously or upon encounter with
certain rates.
Since particle numbers are changed by integer numbers $\{ n_\alpha \}$ in these 
stochastic processes, the corresponding loss and gain terms in the associated 
master equations can be expressed through the action of bosonic creation and 
annihilation operators on a Fock space state vector $| \{ n_\alpha \} \rangle$ 
that contains a list of all species' particle occupations \cite{Doi76, 
Grassberger80, Mattis98}.
One may then utilize a coherent-state basis for the resulting non-Hermitean 
many-body problem to construct an equivalent Doi--Peliti path-integral 
representation \cite{Peliti85, Tauber05, Andreanov06, Cardy08, Tauber14}.
If the occupation numbers are restricted to just $n_\alpha = 0$ or $1$, the 
stochastic reactions can in contrast be represented through spin-$1/2$ operators.
This mapping is especially fruitful in one dimension, where mathematical tools
such as the Bethe ansatz developed for quantum spin chains can be applied to the
ensuing non-Hermitean spin Hamiltonians \cite{Alcaraz94, Henkel97, Schutz01, 
Stinchcombe01}.

For at most binary reactions, the path integral action can be cast in the form 
\ref{janded}. albeit with complex fields $\psi^\alpha({\vec x},t)$; 
$F^\alpha[\psi]$ then represents the reaction functional as familiar from the 
mass action expression or chemical rate equation in well-mixed systems. 
One may therefore write down an effective coarse-grained Langevin description,
with usually multiplicative internal reaction noise encoded in the functional 
$L^\alpha[\psi]$.
Let us consider the simplest scenario, namely diffusing particles of species $A$
subject to the annihilation processes $k \, A \to l \, A$ where
$0 \leq l < k \leq 2$.
With continuum diffusion and reaction rates $D$ and $\lambda$, the ensuing
stochastic partial differential equation becomes \cite{Peliti86, Lee94, Tauber05,
Tauber14}
\begin{equation}  
  \frac{\partial \psi({\vec x},t)}{\partial t} = D \, {\vec \nabla}^2 \,
  \psi({\vec x},t) - (k - l) \, \lambda \, \psi({\vec x},t)^k 
  + \zeta({\vec x},t) \ ,
\label{annlan}
\end{equation}
with $\langle \zeta({\vec x},t) \rangle = 0$ and the formal noise correlator
\begin{equation}
  \left\langle \zeta({\vec x},t) \, \zeta({\vec x\,}',t') \right\rangle 
  = - 2 \left[ k (k - 1) - l (l - 1) \right] \lambda \, \psi({\vec x},t)^k \, 
  \delta({\vec x} - {\vec x\,}') \, \delta(t - t') \ .
\label{annnoi}
\end{equation}

\begin{figure}[h] \vskip -0.3cm
\includegraphics[width=6.5cm]{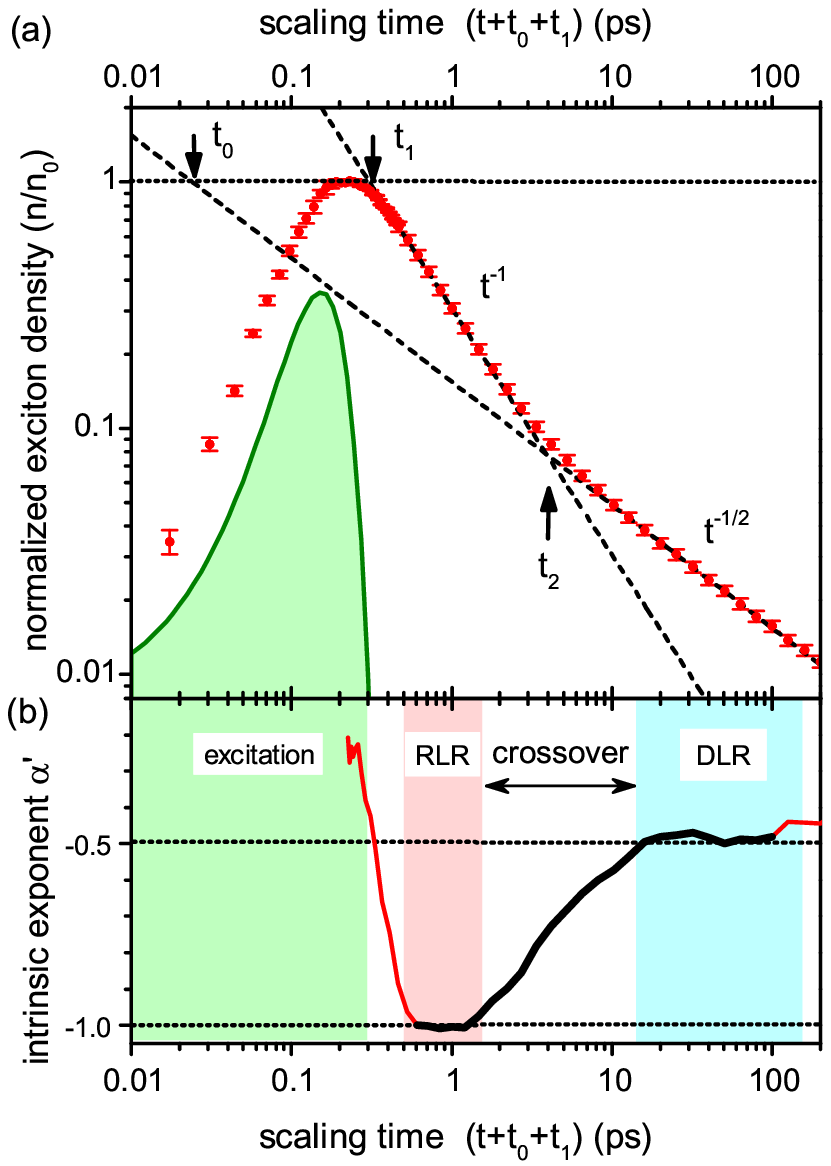} \vskip -0.2cm
\caption{Experimental exciton recombination data in single-walled carbon 
   nanotubes that clearly demonstrate the crossover from reaction- to 
   diffusion-limited power law decay: (a) normalized exciton density as
   function of time; (b) the intrinsic effective decay exponent displays
   plateaus at the reaction-limited value $-1$ and renormalized 
   diffusion-limited $-1/2$ (Figures reproduced with permission from 
   Ref.~\cite{Allam13}, DOI: 10.1103/PhysRevLett.111.197401;
   copyright 2013 by the American Physical Society).}
\label{fig3}
\end{figure}
For $k = 1$, i.e., spontaneous death $A \to \emptyset$ at rate $\lambda$, the 
stochastic noise vanishes, and the mean particle density of course decays 
exponentially, 
$a(t) = \langle \psi({\vec x},t) \rangle = a(0) \, e^{- \lambda \, t}$.
In contrast, the negative sign in \ref{annnoi}. for $k = 2$ indicates emerging 
spatial anti-correlations, as the pair annihilation reactions quickly remove 
near-by particles.
Neglecting temporal fluctuations and spatial correlations, i.e., omitting the
noise and diffusive spreading in \ref{annlan}., the ensuing mean-field rate
equation for the particle density is for $k \geq 2$ solved by $a(t) = 
 \left[ a(0)^{1 - k} + (k - l) (k - 1) \, \lambda \, t \right]^{- 1 / (k - 1)}$,
which asymptotically yields a power law decay that is independent of the initial 
value $a(0)$.
Since the field $[\psi({\vec x},t)] = \mu^d$ should scale like a density, one
obtains the scaling dimension $[\lambda] = \mu^{2 - (k - 1) d}$ for the $k$-th
order annihilation rate.
Consequently the critical dimension for stochastic annihilation is 
$d_c(k) = 2 / (k - 1)$; for $k \geq 4$ therefore, fluctuations do not markedly
modify the mean-field reaction-limited decay law in any physical dimension $d$.
For pair ($k = 2$) annihilation $A + A \to \emptyset$ or coagulation 
$A + A \to A$, the reactions generate spatial depletion zones of typical length 
$L(t)$ that need to be traversed diffusively by other particles before further 
annihilations can ensue.
In this diffusion-limited regime, therefore $L(t) \sim (D t)^{1/2}$, whence 
$a(t) \sim L(t)^{-d} \sim (D \, t)^{- d / 2}$, which represents a slower decay 
than the reaction-limited $a(t) \sim (\lambda \, t)^{-1}$ in dimensions 
$d \leq 2$ \cite{Peliti86, Lee94, Tauber05, Krapivsky10}.
Precisely at the critical dimension $d_c(2) = 2$, and $d_c(3) = 1$ for triplet
annihilation, one obtains logarithmic corrections: 
$a(t) \sim \left[ (D \, t)^{-1} \, \ln (D \, t) \right]^{1 / (k - 1)}$.
Experimentally, the anomalous diffusion-limited power law decay in one dimension
has been verified in the exciton recombination kinetics in molecular wires
\cite{Kopelman88}, TMMC polymer chains \cite{Kroon93}, and carbon nanotubes
\cite{Russo06}; the convincing data from Ref.~\cite{Allam13} shown in 
Fig.~\ref{fig3} provide clear evidence for the crossover from the reaction- to 
diffusion-limited regimes.

Additional and quite different physical mechanisms govern the emerging spatial
correlations in two-species pair annihilation $A + B \to \emptyset$, where 
crucially the particle number difference of the two species remains conserved
under the time evolution \cite{Toussaint83}.
With equal diffusivities for both species, the associated stochastic 
differential equations read \cite{Lee95, Tauber14}
\begin{eqnarray}
  &&\frac{\partial \psi({\vec x},t)}{\partial t} = D {\vec \nabla}^2 
  \psi({\vec x},t) - \lambda \, \psi({\vec x},t) \, \varphi({\vec x},t) 
  + \zeta({\vec x},t) \ , \nonumber \\
  &&\frac{\partial \varphi({\vec x},t)}{\partial t} = D {\vec \nabla}^2 
  \varphi({\vec x},t) - \lambda \, \psi({\vec x},t) \, \varphi({\vec x},t) 
  + \eta({\vec x},t) \ ,
\label{abalan}
\end{eqnarray}
with $\langle \zeta({\vec x},t) \rangle = 0 = \langle \eta({\vec x},t) \rangle$
and the noise (cross-)correlations
\begin{eqnarray}
  &&\langle \zeta({\vec x},t) \, \zeta({\vec x\,}',t') \rangle = 0 
  = \langle \eta({\vec x},t) \, \eta({\vec x\,}',t') \rangle \ , \nonumber \\
  &&\langle \zeta({\vec x},t) \, \eta({\vec x\,}',t') \rangle = - \lambda \, 
  \psi({\vec x},t) \, \varphi({\vec x},t) \, \delta({\vec x} - {\vec x\,}') \,
  \delta(t - t') \ .
\label{abanoi}
\end{eqnarray}
For unequal initial densities $a(0) - b(0) = a(\infty) > 0$, only the majority
species $A$ will survive as $t \to \infty$, and in dimensions $d > 2$, the 
approach to the stationary values is exponential: 
$a(t) - a(\infty) \sim b(t) \sim e^{- \lambda \, a(\infty) \, t}$.
In low dimensions $d < 2$, depletion zone anti-correlations induce a slower,
stretched-exponential decay, 
$\ln \left[ a(t) - a(\infty) \right] \sim \ln b(t) \sim - (D t)^{d / 2}$, 
whereas at $d_c = 2$ one obtains 
$\ln \left[ a(t) - a(\infty) \right] \sim \ln b(t) \sim - D t / \ln (D t)$.
In stark contrast, for equal initial densities $a(0) = b(0)$, the mean-field
rate equations predict $a(t) \sim b(t) \sim (\lambda \, t)^{-1}$ as for 
single-species pair annihilation.
This becomes modified by spatial particle species segregation in dimensions
$d \leq d_s = 4$; exploiting that 
$c({\vec x},t) = \psi({\vec x},t) - \varphi({\vec x},t)$ becomes a purely
diffusive mode, straightforward analysis shows that the local density excess
decays slowly, $|c(x,t)| = 2 \sqrt{a(0) / \pi} \, (8 \pi \, D \, t)^{- d / 4}$,
whence also $a(t) \sim b(t) \sim (D \, t)^{- d / 4}$.
The annihilation reactions become confined to narrow reaction zones that 
separate segregated and inert $A$ and $B$ domains.
The reaction zone width scales as $w(t) \sim (D \, t)^\Lambda$, where 
$\Lambda = 1/3$ in the mean-field approximation, while 
$\Lambda = (d + 2) / 4 \, (d + 1)$ for $d \leq d_c = 2$.
The three-dimensional non-classical decay $\sim t^{- 3/4}$ has been observed 
experimentally \cite{Monson04}.

Note that the highly correlated alternating initial arrangement 
$\ldots ABABABABAB \ldots$ is preserved by the reactions $A + B \to \emptyset$ 
in one dimension; hence the distinction between $A$ and $B$ particles becomes 
meaningless, and their densities indeed satisfy the single-species $t^{- 1/2}$ 
pair annihilation power law.
One may in fact fully analyze the $q$-species pair annihilations 
$A_i + A_j \to \emptyset$, $1 \leq i < j \leq q$, with equal initial densities 
$a_i(0)$ as well as uniform diffusion and reaction rates \cite{Deloubriere02,
Deloubriere04, Hilhorst04}.
Indeed, for more than two species ($q > 2$), there exists no conservation law 
in the stochastic kinetics, and species segregation results for 
$d < d_s(q) = 4 / (q - 1)$.
For $d \geq 2$, one therefore obtains the single-species pair annihilation 
decay laws.
On one dimension, each species' density obeys  
$a_i(t) \sim t^{- \alpha(q)} + C \, t^{- 1/2}$, i.e., a leading slow power law
decay with $\alpha(q) = (q - 1) / 2 q$ induced by segregation effects, 
accompanied with a subleading term caused by depletion zones.   
For the reaction front width, one finds the scaling behavior 
$w(t) \sim t^{\Lambda(q)}$ with $\Lambda(q) = (2 q - 1) / 4 q$.

\subsection{Active-to-Absorbing State Phase Transitions}

If in addition to particle decay $A \to \emptyset$ and annihilations, e.g.,
$A + A \to A$, one allows for competing offspring production through branching 
processes such as $A \to A + A$, the overall particle density may at long times
either reach a finite mean value, representing an active state, or vanish.
In situations where the presence of particles is required for any generation of
further offspring, i.e., in the absence of spontaneous particle production 
$\emptyset \to A$, the latter inactive phase is also absorbing: 
Once reached, there are no stochastic processes available that would allow the
system to escape this empty state \cite{Hinrichsen00, Odor04, Henkel08}.
Although we have just formulated the active-to-absorbing transition scenario in
terms of chemical reactions, the particle-like excitations could also be domain
walls that coalesce or annihilate, or other effective coarse-grained degrees of
freedom whose kinetics is captured by these stochastic interactions. 
More generally, one may consider spreading activity fronts or an infectious 
epidemic which on a lattice would be represented by discrete entities $A$.

Indeed, heuristic considerations permit a straightforward phenomenological 
continuum description of the so-called simple epidemic process or a spreading 
epidemic with recovery \cite{Murray02}:
Assuming diffusive spreading with diffusion constant $D$, and strictly local 
infections of a homogeneous susceptible medium, we write down the Langevin 
equation \cite{Janssen05, Tauber14}
\begin{equation}
  \frac{\partial S({\vec x},t)}{\partial t} = D \, {\vec \nabla}^2 S({\vec x},t) 
  - R[S({\vec x},t)] \, S({\vec x},t) + \zeta({\vec x},t) \ ,
\label{seplan}
\end{equation}
where $R[S]$ denotes an appropriate reaction functional with finite limit as
the active or infected density $S \to 0$. 
In addition, the and multiplicative stochastic noise with vanishing mean
$\langle \zeta({\vec x},t) \rangle = 0$ and correlator
\begin{equation}
  \left\langle \zeta({\vec x},t) \, \zeta({\vec x\,}',t') \right\rangle 
  = 2 L[S({\vec x},t)] \, \delta({\vec x} - {\vec x\,}') \, \delta(t - t')
\label{sepnoi}
\end{equation} 
that represents all other fast degrees of freedom and internal reaction noise 
must also satisfy the absorbing-state condition, whence $L[S] \to 0$ as 
$S \to 0$.
In the vicinity of the extinction threshold, we may approximate both reaction 
and noise functionals in the spirit of a Landau expansion:
$R[S] = D \, r + u \, S \, + \ldots$ and $L[S] = v \, S + \ldots$.
After straightforward rescaling, we may set $v = u$; dimensional analysis with
$[S({\vec x},t)] = \mu^{d / 2}$ and $[D] = \mu^0$ then yields $[r] = \mu^2$ and
$[u] = \mu^{2 - d /2}$.
Thus $d_c = 4$ is the upper critical dimension for the absorbing-state phase
transition, and omitting higher-order terms as well as powers of gradients of 
the activity field become a-posteriori justified, since all such additional 
contributions turn out to be irrelevant in the renormalization group sense.
Neglecting the multiplicative noise, \ref{seplan}. reduces to the 
deterministic Fisher--Kolmogorov reaction-diffusion equation \cite{Haken83, 
Murray02}.
The phase transition to the absorbing state occurs at $r = 0$; in the active
state for $r < 0$ one has $\phi(\infty) = D \, |r| / u$ as $t \to \infty$ or
$\beta = 1$; diffusive propagation implies $z = 2$, and the mean-field
critical correlation exponents are $\eta = 0$ and $\nu = 1/2$.
Precisely at the extinction threshold, the mean density decays algebraically:
$\phi(t) = \langle S({\vec x},t) \rangle \sim t^{- \alpha}$ with 
$\alpha = \beta / z \, \nu$.

\begin{figure}[h] \vskip -0.6cm
\centerline{\includegraphics[width=6.6cm]{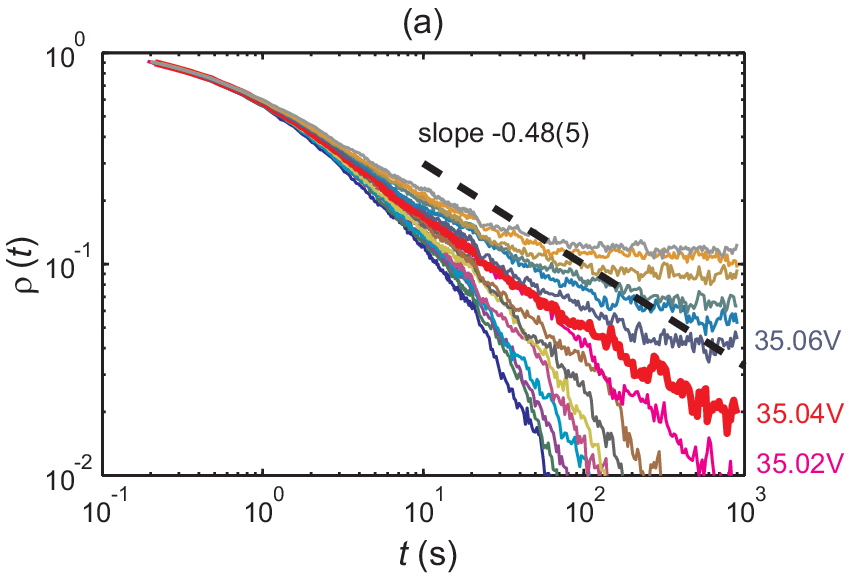} \hskip -6.8cm
\includegraphics[width=6cm]{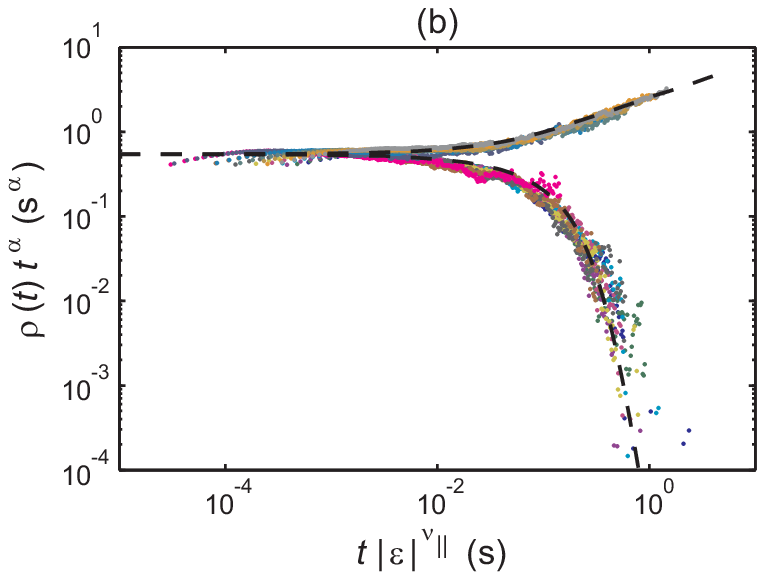}} \vskip 0.2cm
\caption{Evidence for directed percolation critical behavior in effectively
   two-dimensional turbulent liquid crystals: 
   (a) active density decay in the vicinity of the critical voltage;
   (b) scaling collapse with critical exponents $\alpha \approx 0.48$ and 
   $\nu_\parallel \approx 1.29$; the thick dashed lines indicate the scaling 
   functions in the active (upper curve) and absorbing (lower curve) phases 
   obtained from contact process Monte Carlo simulations (Figures reproduced 
   with permission from Ref.~\cite{Takeuchi09}, DOI: 10.1103/PhysRevE.80.051116; 
   copyright 2009 by The American Physical Society).}
\label{fig4}
\end{figure}

The Janssen--De~Dominicis response functional \ref{janded}. for the stochastic 
differentical equation \ref{seplan}. with multiplicative noise \ref{sepnoi}. and 
just the leading and relevant terms retained in the functional expansions is 
known as Reggeon field theory \cite{Moshe78}; it in turn represents the 
effective action for the universal scaling properties of directed percolation 
\cite{Cardy80}: 
Here, the time coordinate maps onto a singled-out spatial direction, and the
decay, annihilation, and reproduction processes respectively correspond to 
terminal, coalescing, or splitting branches in the ensuing percolation clusters
\cite{Obukhov80}.
Its characteristic symmetry feature is rapidity invariance, which entails time
inversion and exchange of the dynamical fields:
$S({\vec x},t) \leftrightarrow - {\widetilde S}({\vec x}, - t)$. 
These considerations establish the Janssen--Grassberger conjecture, which states
that the asymptotic critical features for continuous non-equilibrium phase 
transitions from active to inactive, absorbing states that are described by a  
scalar order parameter field and governed by Markovian stochastic dynamics that 
is decoupled from any other slow variables and not subject to the influence of 
quenched disorder, are generically captured by the universality class of 
directed percolation \cite{Janssen81, Grassberger82}.
The associated critical and aging scaling exponents are numerically known to 
high accuracy in all dimensions $d < d_c = 4$ \cite{Henkel08}, and can be 
systematically computed in a $d = 4 - \epsilon$ expansion by means of the 
perturbative dynamical renormalization group \cite{Janssen05, Tauber14}.
Experimentally, directed percolation scaling has been observed in intermittent
ferrofluidic spikes \cite{Rupp03}, and unambiguously at the transition between 
different turbulent states of electro-hydrodynamic convection in thin
nematic liquid crystals \cite{Takeuchi07, Takeuchi09}.
Figure~\ref{fig4} shows the active density decay near criticality and data
collapse obtained with directed percolation scaling exponents.

While coupling to other slow modes or quenched randomness in the percolation
threshold may invalidate directed percolation scaling \cite{Henkel08}, this 
universality class still pertains for an extension of active-to-absorbing state 
transitions to multiple particle species \cite{Janssen97, Janssen01}, except for 
special multi-critical points in parameter space \cite{Tauber98, Goldschmidt99}.
For example, in spatially extended stochastic Lotka--Volterra predator-prey 
models \cite{Haken83, Murray02}, local restrictions of the prey carrying 
capacity induces a predator extinction threshold. 
This continuous transition displays directed-percolation critical exponents, and
its effective Doi--Peliti action can be mapped to Reggeon field theory
\cite{Mobilia07, Tauber12}.
Numerical simulations confirm that population extinctions are generically 
described by the directed percolation scaling laws, including the associated
critical aging exponents \cite{Mobilia07, Chen16}.
Another remarkable analogy addresses the onset of shear turbulence in pipe flow,
which displays very similar spatio-temporal phenomena as predator-prey kinetics,
including spreading activity fronts; its threshold properties have hence been 
argued to belong to the directed-percolation universality class as well 
\cite{Shih15}.

In general epidemic processes \cite{Murray02}, infected sites or individuals 
never return to a susceptible state, but remain infectious to their neighbors.
This induces persistent temporal memory into the non-linearity in \ref{seplan}.
of the form $D \, u \, S({\vec x},t) \int_{- \infty}^t S({\vec x},t') \, dt'$
\cite{Grassberger83, Cardy85, Janssen85}.
This enhances the influence of correlations and shifts the upper critical 
dimension to $d_c = 6$; interestingly, the ensuing quasi-static scaling 
properties near the infection threshold are identical to those for static 
isotropic percolation clusters \cite{Stauffer94}.
Active-to-absorbing state transitions also occur in branching and annihilating 
random walks wherein diffusing particles are subject to the competing reactions 
$A + A \to \emptyset$ and $A \to (m + 1) \, A$ with integer $m$ \cite{Tauber96, 
Cardy98}.
Here, the situations of odd or even $m$ are crucially different:
For odd $m = 1, 3, \ldots$, fluctuations that combine both these processes 
generate the spontaneous decay $A \to \emptyset$, and the phase transition is 
governed by directed-percolation scaling exponents.
However, that is not possible for branching reactions with even 
$m = 2, 4, \ldots$, as in this case the particle number parity is a locally 
conserved quantity. 
In fact, the entire absorbing state of this parity-conserving universality class 
is characterized by the pure binary-annihilation decay laws, and hence by
scale-invariant dynamical correlations.
It actually constitutes a special case for a generalized voter model universality
class that obeys a $Z_2$ inversion symmetry \cite{Chate05}.
Replacing first-order branching with the binary reaction $A + A \to (m + 2) \, A$
yields the pair contact process with diffusion. 
If arbitrarily many particles are allowed per lattice site, it displays a 
strongly discontinuous non-equilibrium phase transition \cite{Howard97}; in
contrast, if site occupation restrictions are implemented, its absorbing state
transition becomes continuous, but owing to exceedingly long crossovers, its
precise asymptotic scaling properties are still controversial \cite{Henkel04}.
Despite extensive efforts, e.g. \cite{Elgart06}, a full classification of 
active-to-absorbing phase transitions that incorporate triplet or higher-order
reactions has remained elusive to date.

\section{CONCLUSION AND OUTLOOK}

Below, I list the main conclusions of this brief (and by necessity incomplete) 
introductory overview on continuous phase transitions and the emergence of 
dynamical scaling in non-equilibrium systems, and provide my personal outlook to 
crucial future research avenues.

\begin{summary}[SUMMARY POINTS]
\begin{enumerate}
\item In thermal equilibrium, universality classes for critical dynamics are 
	defined through the symmetries of the order parameter, conservation laws, and
	possible couplings of the order parameter to other slow conserved modes.
\item Dynamical scaling concepts have been successfully extended to quantum
	phase transitions and open systems far from equilibrium.
\item Aging scaling emerges in the non-equilibrium relaxation kinetics of systems 
    with either exceedingly long relaxation times or slow algebraic decay.
\item Generic scale invariance is a quite prevalent feature of non-equilibrium
	stationary states in driven systems.
\item Driven-diffusive systems are characterized by anisotropic	scaling laws.
\item The Kardar--Parisi--Zhang model and variants describe universal scaling 
	features and a non-equilibrium roughening transition of driven interfaces or
	growing surfaces.
\item In diffusion-limited reacting particle systems at low dimensions, strong 
	temporal fluctuations and spatial correlations invalidate standard mean-field
	rate equation descriptions that utilize mass-action type factorizations.
\item Active-to-absorbing state transitions are generically captured by the 
	universal scaling properties of critical directed percolation.
\end{enumerate}
\end{summary}

\begin{issues}[FUTURE ISSUES]
\begin{enumerate}
\item Reproducible and quantitative experimental verification based on top-quality
	data of the theoretically and numerically established prominent dynamical 
	universality classes out of equilibrium currently only exists for a few 
	important cases (e.g., diffusion-limited annihilation, Kardar--Parisi--Zhang 
	scaling, directed percolation).
\item A full classification of all possible non-equilibrium universality classes 
	remains elusive to date, owing to the increasing complexity as additional 
	relevant degress of freedom or reacting particle or ecological species are
	taken into consideration.
\item The emergence of improved quantitative data in biological systems ranging 
	from biochemical reactions via collective behavior of micro-organisms 
	to macroscopic pattern formation and population extinctions in eco-systems 
	promises to become an ever more fertile ground for the application of scaling
	tools from statistical physics.
\end{enumerate}
\end{issues}

\section*{DISCLOSURE STATEMENT}
The author is not aware of any affiliations, memberships, funding, or financial 
holdings that might be perceived as affecting the objectivity of this review. 

\section*{ACKNOWLEDGMENTS}
The author gratefully acknowledges his many invaluable research collaborators and
students, as well as financial support by the U.S. Department of Energy, Office 
of Basic Energy Sciences, Division of Materials Sciences and Engineering under 
Award DE-FG02-09ER46613.

\end{document}